\documentclass[aps,showpacs,superscriptaddress,twocolumn]{revtex4}

\usepackage{amsfonts}
\usepackage{amsmath}
\usepackage{amsthm}
\usepackage{amscd}
\usepackage{amssymb}
\usepackage{subfigure}
\usepackage{amsxtra}
\usepackage{bm}           
\usepackage{bbm}
\usepackage{exscale}
\usepackage{epic,rotating}
\usepackage{graphics,color,dsfont}
\usepackage{latexsym}
\usepackage{enumerate}
\usepackage{dcolumn}      
\usepackage{pstricks,pst-grad,fancybox,graphics}
\usepackage[dvips]{epsfig}
\usepackage[crop=off]{auto-pst-pdf} 

\def\bi#1\ei {\begin{itemize}#1\end{itemize}}
\def\bn#1\en {\begin{enumerate}#1\end{enumerate}}
\def\bea#1\eea {\begin{align}#1\end{align}}
\def\bean#1\eean {\begin{align*}#1\end{align*}}
\def\ben#1\een {\begin{equation*}#1\end{equation*}}
\def\be#1\ee {\begin{equation}#1\end{equation}}
\def\bes#1\ees {\begin{equation}\begin{split}#1\end{split}\end{equation}}
\def\bear#1\eear {\begin{eqnarray}#1\end{eqnarray}}
\def\bear#1\eear {\begin{eqnarray*}#1\end{eqnarray*}}

\newcommand{\vek}[1]{\ensuremath{\mbox{\bf vec}\left(#1\right)}}
\newcommand{\nullsp}[1]{\ensuremath{\mbox{\bf Null}\left(#1\right)}}

\newcommand{\bra}[1]{\ensuremath{\langle#1|}}
\newcommand{\ket}[1]{\ensuremath{|#1\rangle}}
\newcommand{\braket}[2]{\ensuremath{\langle #1|#2\rangle}}
\newcommand{\ketbra}[1]{\ensuremath{| #1 \rangle \langle #1 |}}
\newcommand{\ketbraa}[1]{\ensuremath{| #1 \rangle_{\alpha} \langle #1 |}}
\newcommand{\ketbraaa}[3]{\ensuremath{| #1 \rangle_{#2}^{#3} \langle #1 |}}

\newcommand{\proj}[1]{\ketbra{#1}}
\newcommand{\eins}{\ensuremath{\mathbbm 1}}
\renewcommand{\qed}{\ensuremath{\hfill \blacksquare}\medskip}
\newcommand{\HH}{\ensuremath{\mathcal{H}}}
\newcommand{\OO}{\ensuremath{\mathcal{O}}}
\newcommand{\BB}{\ensuremath{\mathcal{B}}}

\newcommand{\MM}{\ensuremath{\mathcal{M}}}
\newcommand{\PP}{\ensuremath{\mathcal{P}}}
\renewcommand{\vr}{\ensuremath{\rho}}
\newcommand{\tr}[1]{\ensuremath{\mbox{Tr}\left( #1 \right)}}

\newcommand{\FF}{\ensuremath{\mbox{{\bf F}}}}
\newcommand{\CC}{\ensuremath{\mathbbm{C}}}

\newcommand{\halbe}{\ensuremath{\frac{1}{2}}}
\newcommand{\ftil}{\ensuremath{\tilde{F}}}

\newtheorem{thm}{Theorem}
\newtheorem{lem}[thm]{Lemma}
\newtheorem{corol}[thm]{Corollary}
\newtheorem{prop}[thm]{Proposition}
\newtheorem{defn}[thm]{Definition}

\newtheorem{rem}[thm]{Remark}
\newtheorem{obs}[thm]{Observation}

\begin{document}

\title{Squashing model for detectors and applications to quantum key distribution protocols}

\author{O. Gittsovich}
\affiliation{Institute for Quantum Computing \&
Department of Physics and Astronomy,
University of Waterloo, Waterloo ON Canada
N2L 3G1}

\author{N. J. Beaudry}
\affiliation{Institute for Quantum Computing \&
Department of Physics and Astronomy,
University of Waterloo, Waterloo ON Canada
N2L 3G1}

\affiliation{Institute for Theoretical Physics, ETH Z\"urich, 8093 Z\"urich, Switzerland}

\author{V. Narasimhachar}
\affiliation{Institute for Quantum Computing \&
Department of Physics and Astronomy,
University of Waterloo, Waterloo ON Canada
N2L 3G1}

\author{R. Romero Alvarez}
\affiliation{Institute for Quantum Computing \&
Department of Physics and Astronomy,
University of Waterloo, Waterloo ON Canada
N2L 3G1}

\affiliation{Department of Physics, University of Toronto,
Toronto ON Canada M5S 1A7}

\author{T. Moroder}
\affiliation{Institute for Quantum Computing \&
Department of Physics and Astronomy,
University of Waterloo, Waterloo ON Canada
N2L 3G1}

\affiliation{
Naturwissenschaftlich-Technische Fakult\"at,
Universit\"at Siegen, Walter-Flex-Stra{\ss}e 3, D-57068 Siegen, Germany
}

\author{N. L\"utkenhaus}
\affiliation{Institute for Quantum Computing \&
Department of Physics and Astronomy,
University of Waterloo, Waterloo ON Canada
N2L 3G1}

\begin{abstract}
We develop a framework that allows a description of measurements in Hilbert spaces that are smaller than their
natural representation. This description, which we call a ``squashing model'',
consists of a squashing map that maps the input states
of the measurement from the original Hilbert space to the smaller one, followed by a targeted prescribed
measurement on the smaller Hilbert space. This framework has applications in quantum key distribution,
but also in other cryptographic tasks, as it greatly simplifies the theoretical analysis
under adversarial conditions.

\end{abstract}


\date{\today}

\maketitle

\section{Introduction}
Measurements are an essential part of quantum mechanics. In quantum communication, among other fields, various measurements are used to extract information from signals. In quantum cryptographic contexts measurement results often allow the inference of  how third parties are correlated with the obtained data. Usually, the quantum advantage of these communication protocols is demonstrated in theoretical protocols utilizing abstract qubit systems, or other low dimensional systems. However, in physical realizations of quantum communication protocols, no qubit systems are available; instead, one resorts to optical implementations where the signals and measurements are described on infinite-dimensional Hilbert spaces corresponding to optical modes.

In the realm of quantum optics experiments, we are used to the idea of approximating these infinite-dimensional systems easily by lower dimensional descriptions, e.g.~describing parametric down-conversion experiments only on the level of vacuum and single photon pairs. We can do this because we can handle the approximations well on a theoretical level such that theoretical predictions and experimental verifications coincide with high precision.

In quantum cryptographic situations, such as quantum key distribution (QKD) or quantum coin tossing
\cite{bennett84a,bennett92d,berlin09a,berlin11a} this is not good enough. In such contexts, we would have to account for the information that an arbitrary third party could gain about our measurement data. Since experimental verification of third-party information is not possible, we need to be able to provide rigorous bounds on such compromised information. One possibility is to do full calculations in
the infinite-dimensional Hilbert spaces \cite{nl99a,nl00a}. Often this is technically challenging.
The other possibility is to do truncations to finite-dimensional subspaces. These cannot be in the form of approximations, but as truncations that also hold under adversarial
conditions. Again, there are two possibilities. The traditional way would be to provide exact
bounds on the effect of truncations and to extend the theoretical qubit analysis to accommodate
the effects of the truncation. This approach has been followed, for example, in Ref.~\cite{koashisub09a}
in the context of a specific application, while a more general framework of this approach has recently
been formulated in Ref.~\cite{fung11a}.  Here we show a second way, which was already postulated in
Ref.~\cite{gottesman04a}, where the term ``squashing'' was coined for this approach. The squashing method performs a truncation of the Hilbert space in such a
way that provides a direct link between the optical implementation and the abstract low dimensional protocol,
without the necessity to amend the theoretical analysis in the truncated Hilbert space. In the
context of QKD, this approach means that for a generic QKD protocol with a BB84 \cite{bennett84a}
polarization encoding we can assume without loss of generality that single photons enter the
detection device of the receiver.

Thus, our approach allows a truncation of high dimensional Hilbert spaces to some low dimensional target space that also holds under adversarial conditions, as they occur in cryptographic contexts. We build on our earlier work \cite{beaudry08a} that gave a well-defined notion of a squashing map that allows us to clarify the role of the squashing assumption. Note that Tsurumaru and Tamaki \cite{tsurumaru08a,tsurumaru10a} independently investigated squashing models.

\begin{figure}
\includegraphics[width=.99\columnwidth]{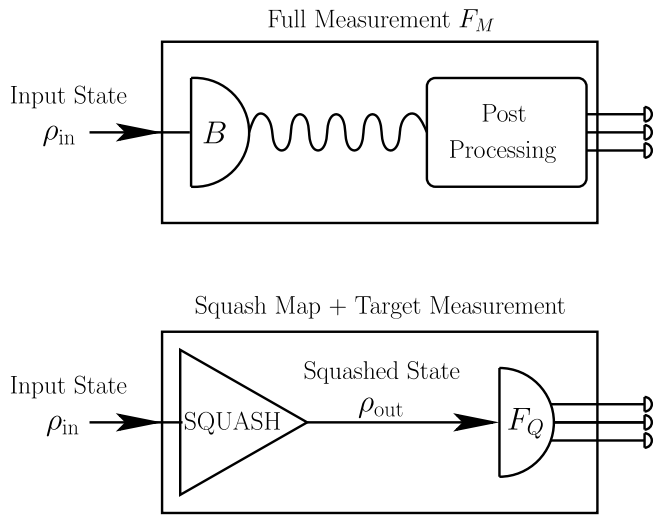}
\caption{The full measurement $F_M$ (above) has a general optical
input $\vr_{in}$, which is first measured by a receiver's measurement
device $B$, followed by classical postprocessing. The
squashed measurement (below) has the same general optical
input $\vr_{in}$, which is then squashed by a map SQUASH to a smaller
Hilbert space, followed by a fixed physical measurement $F_Q$.
It is required that both of these measurements produce the
same output statistics for all $\vr_{in}$. \label{squashidea}}
\end{figure}

A rough idea of what a squashing model does is represented in Fig.~\ref{squashidea}. Each physical measurement device $B$ provides
some basic distinguishable events for an input state $\vr_{\textnormal{in}}$, which usually has its support on a high
dimensional Hilbert
space. All possible events that can be triggered by the states in the high dimensional Hilbert space will formally
correspond to a positive operator-valued measure (POVM) $F_B$.
If one would feed the measurement device with states from some low dimensional Hilbert space,
then typically the number of possible events will be smaller.
We refer the corresponding POVM as to target POVM and denote it by $F_Q$.
Often, however, the events produced by the states from the high dimensional and low dimensional Hilbert spaces can be related by
a classical postprocessing,
which is applied to the basic events. A typical example in QKD is a processing of
double clicks occurring in the BB84 protocol.
The basic events after a particular (classical) postprocessing form a coarse-grained list of events, which then are described by a POVM $F_M$. We refer to the combination of basic events and postprocessing as the {\it full measurement}.
This classical postprocessing will be an essential tool in making squashing models work.
A squashing model provides an equivalent but simplified description the full measurement $F_M$ on the high dimensional Hilbert space in terms of the measurement $F_Q$ on the low dimensional Hilbert space.
The map SQUASH or squashing map (see Fig. \ref{squashidea}) provides a direct link between the measurements on the high
dimensional and the truncated Hilbert space. Formally, the squashing map takes a state $\vr_{\textnormal{in}}$ as input
and outputs a `squashed' state $\vr_{\textnormal{out}}$ on the truncated Hilbert space. The squashing model is an
equivalent description of the
full measurement $F_M$ in terms of the squashing map and a target measurement $F_Q$ on a squashed state $\vr_{\textnormal{out}}$.
All elements (basic measurement $F_B$, target measurement $F_Q$, and classical postprocessing) need to be specified in order to form a well
posed question for the existence of a squashing model. A typical choice for the  target measurement will be the restriction of the
full measurement to a single-photon input, though our framework is not limited by this particular choice.

In this article we first extend the formalism introduced in \cite{beaudry08a} and provide a rigorous framework for
how to find a squashing model for a particular measurement device and which general steps can be
used in order to simplify the analysis. For
example we
show how to enforce the existence of squashing maps by choosing the postprocessing that introduce additional noise.
Second, we review previous
results involving the squashing model for the measurement devices which are used in optical implementations
of the BB84 \cite{beaudry08a} and six-state \cite{beaudry08a,tsurumaru08a} QKD protocols with an active detection scheme.
Third, we discuss several generalizations of these measurement devices and find new squashing
models for the corresponding devices.
For instance, we present a squashing model for a generalization of the qubit measurement devices with the passive
detection scheme to qudit measurement devices \cite{bechmann99a,cerf02a,durt_security_2004} and prove that there exists a
squashing model for this generalized device. We also
consider squashing models for the measurements that accept different temporal modes
and are employed in the phase encoded BB84 (PEBB84) protocol.

This paper is organized as follows. The first part (Sections \ref{notsndefs} - \ref{linoptimpl}) is devoted
to the general framework and discussion of the general properties of the examples presented in the
second part (Sections \ref{ssection:BB84active} - \ref{Application}). Some technical details relevant for our
investigations are given in the appendix.

In Section \ref{notsndefs} we fix the notation and define the
quantities that will be frequently used. We will define the squashing model as well. In Section \ref{genstrat}
we present general strategies for finding a squashing map for a general measurement device and discuss
possible issues such as non-positivity of the squashing map.
Subsequently, in Section \ref{linoptimpl}, we consider common properties of typical linear optical measurement devices,
which simplify the construction of the squashing model. In particular, we discuss how the usage
of threshold detectors helps to truncate infinite-dimensional Hilbert space of an optical mode and
the consequences it has for the application of the general framework to concrete examples.

In Sections \ref{ssection:BB84active}-\ref{section:qubitextensions}
we apply the presented theory in order to construct squashing models for several optical measurement devices.
These include: active measurement devices from the BB84 QKD protocol (Section \ref{ssection:BB84active}), from the six-state protocol (Section \ref{ssection:6stateact}), biased active and passive measurements from
the BB84 protocol, and passive measurement from the six-state QKD protocol (Section \ref{section:qubitextensions}).
In Section \ref{rubenswork} we present a new squashing
model for measurement devices that can be used in the optical implementation of qudit QKD protocols.
In Section \ref{sectm} we consider squashing models in the time domain, where the incoming state can carry multiple
photons that are distributed over several time modes.
We show that this generalization does not affect the existence of a squashing model. Finally in Section
\ref{Application} we discuss the squashing model for the measurement device from the phase-encoded BB84
(PEBB84) protocol. This refines the security analysis of the corresponding QKD protocol, which was performed
in Refs.~\cite{ferenczi12b,sunohara13suba}.

\section{Notation and statement of the problem}\label{notsndefs}
We first make some preliminary definitions so that we can define a squashing model explicitly.
$\HH_M$ denotes a high dimensional Hilbert space. Basic and full measurements on a state
$\vr_M\in\BB(\HH_M)$ are described by POVMs $F_B$ and $F_M$ respectively. $\HH_Q$ denotes the
low dimensional Hilbert space (i.e.~the target Hilbert space). The measurement on the states $\vr_Q\in\BB(\HH_Q)$ (the target measurement)
is described by the POVM $F_Q$.
Elements of the corresponding POVM are denoted by $F_X^{(i)}$ with $X \in {B,M,Q}$, and the indices $i$ run over the set of outcomes for each of the measurements.
Observed probabilities for measurement outcomes are given by $p_X^{(i)} = \tr{F_X^{(i)} \vr}$.

Moving forward to the formal definition of a squashing model we need to explicitly state
what classical postprocessing means. The classical postprocessing is applied to the basic
measurement outcomes and allows, for example, to combine different outcomes into one (which we call coarse graining).
More precisely, it defines the full measurement by using the basic measurement events such that the POVM
$F_M$ contains the same number of elements as the target POVM $F_Q$.
Otherwise, the problem of finding a squashing model is not well-defined.

Formally, the postprocessing can be
described as a stochastic matrix $\PP$ ($\sum_i\PP_{ij} =1,\forall j$) which acts on the vector
of probabilities of the basic measurement outcomes. The entries of the matrix $\PP_{ij}=p(i|j)$ are given by the
conditional probabilities which describe the redistribution of the outcomes of the POVM $F_B$ with index
$j$ into events of the full measurement POVM $F_M$ with index $i$.

Summarizing the above discussion we have the following
\begin{defn}(Classical postprocessing)\label{CPPdefn}
Let $\vec{p}_{\textnormal{bas}}$ be the vector of the outcome probabilities of the
basic measurement $B$. We say that a classical postprocessing (CPP) scheme is defined
if there exists a stochastic matrix $\PP$ such that
\be
\vec{p} = \PP\vec{p}_{\textnormal{bas}},
\ee
and the number of the outcome probabilities $p_j$ coincides with the number of events provided by
the target POVM $F_Q$.
\end{defn}

It is not hard to see that the postprocessing can be considered as a
linear transformation of the POVM elements
\be
F_M^{(i)}= \sum_j\PP_{ij}F_B^{(j)},
\label{PP}
\ee
where we require that $\vec{p}=\tr{\vr F_M^i}$ describes the vector of outcome probabilities of the
full measurement $F_M$.

In our discussion we will not add a postprocessing step to the target
measurement, as the choice of target measurement is usually motivated by circumstances. A typical example
of this is when a security
proof may exist for a fixed given measurement, which one then typically considers as a target measurement
in the context of the squashing model. As the target measurement is given by a particular POVM,
it may already be a combination of some target measurements and fixed postprocessing of the target events.
An example of this situation will be discussed in Section \ref{Application}, where we construct a
squashing model for the measurement device used in the phase encoded BB84 QKD protocol and will group certain
target measurement events into one (outside clicks).

We will fix the notation and define the CPP and then we will give the formal definition of a squashing model.

\begin{defn}(Squashing model)\label{defsquash}
Let $F_B$ and $F_Q$ be the POVMs that describe outcomes of a measurement performed by a physical
device $B$ on states in high dimensional and low dimensional Hilbert spaces respectively.
Let $\PP$ be a CPP scheme that defines a full measurement POVM $F_M$.
Then we say that there exists a squashing model for the device $B$ and the CPP $\PP$ if there exists a map
$\Lambda_{B}$ such that
\bn
\item
For any state $\vr_M$ the linear constraints
        \be
        \tr{F_{M}^{(i)}\vr_M}=\tr{F_{Q}^{(i)}\Lambda_{B}[\vr_M]}, \forall i\label{probcond}
        \ee
        are satisfied.
\item $\Lambda_{B}$ is a completely positive (CP) map. We call it a ``squashing map''.
\en
\end{defn}
\begin{rem}(Linear constraints on POVM elements)
We introduce the adjoint map $\Lambda_{B}^\dagger$ to find that  Eq.~(\ref{probcond}) implies
        \be
        \tr{F_{M}^{(i)}\vr_M}=\tr{\Lambda_{B}^\dagger [F_{Q}^{(i)}] \vr_M}, \forall i.\label{probcondadjoint}
        \ee
 This has to hold for any state $\vr_M$. Therefore
\be
\Lambda_{B}^{\dagger}[F_{Q}^{(i)}]=F_{M}^{(i)}, \forall i=1,\dots,N_Q,
\label{linconstOperForm}
\ee
which can be seen as linear constraints on the map $\Lambda_B^{\dagger}$
The adjoint map has to satisfy
\bea
\Lambda_{B}^{\dagger}[\eins_Q]&=\eins_M,\label{unital}
\eea
which is the unital property of the adjoint of the squashing map and assures
that $\Lambda_B$ is completely positive and trace preserving (CPTP).
\end{rem}

In order to gain more insight into the formal definition of the squashing model we
make a few more remarks. The definition of the squashing model consists of two essential parts. In order
to provide a squashing model for a given measurement device $B$,
a particular low dimensional Hilbert space must be chosen.
Second, one has to agree on a meaningful postprocessing,
as defined in Definition \ref{CPPdefn}. The classical postprocessing
can be seen as a freedom available to search for a squashing model. That is, the postprocessing
fixes the full measurement and has to satisfy the linear constraints in Eq.~(\ref{probcond}), which
has to be fulfilled by the squashing map $\Lambda_{B}$. In fact, as we will see later on,
for any choice of the POVMs $F_B$ and $F_Q$ there always exists a CPP scheme for which a squashing
model exists. However, as we will also see, such a squashing model may not be meaningful and would correspond to
a very noisy outcome of the measurement.
The squashing map $\Lambda_{B}$ has to be a CPTP map. Therefore its existence, given the constraints, can be investigated by exploiting the Choi-Jamio{\l}kowski isomorphism \cite{jamiolkowski72a,choi75a}. Note that variations of squashing models that require only positive but not completely positive maps have been investigated and utilized in Ref.~\cite{moroder10a}.

\section{General strategy to find a squashing model}\label{genstrat}
The formal definition of the squashing model already provides some
intuition for how to investigate the question of whether a squashing model exists for
a given measurement device. The goal of this section is to provide
a step-by-step strategy to search for a squashing map for any particular
case.

\subsection{Basic and target POVMs}
In our considerations we always assume that the exact physical model of the actual
measurement device is known, so the POVM $F_B$ is fixed. The choice of the target measurement depends on the
choice of the truncated Hilbert space and is always motivated by circumstances. For example, a theoretic analysis of a communication protocol with a specific POVM $F_Q$ might already exist and we would like to link an optical implementation with basic events $F_B$ to this analysis. So the main choice that has to be made to set-up a well defined search for a squashing map is that of the post-processing of the basic events into the full measurement events.

\subsection{Constraints on CPP schemes}\label{ssection:classPP}
We pointed out before that any valid classical postprocessing scheme has to assure that the number of events of the
full and the target measurement coincide. There are further limitations on what types of classical postprocessing that
can lead to a successful squashing map. We note that the set of the target POVM elements $F_{Q}^{(i)}$ may be
linearly dependent which means that there may exist some complex numbers $\alpha_i$, such that
\be
\sum_{i=1}^{N_Q}\alpha_i F_{Q}^{(i)} = 0.\label{eq:targetlindep}
\ee

Each set of POVM elements of the corresponding basic measurement add up to the identity on the
operator space. This also means that the full measurement (including postprocessing) must have the
same linear dependency. This has implications for the postprocessing of the basic events, as
this linear dependence has to be respected by the postprocessing $\PP$ one is
looking for. Due to the linearity of $\Lambda_{B}^{\dagger}$ and the linear constraints in
Eq.~(\ref{linconstOperForm}) we can write
\bea
\sum_{i}\alpha_{i}F_{Q}^{(i)}=0&\Leftrightarrow\sum_{i}\alpha_i F_{M}^{(i)}=0\nonumber\\
&\Leftrightarrow\sum_{i,j}\alpha_{i}\PP_{ij}F_{B}^{(j)}=0.\label{eq:lineardependancegeneral}
\eea
The simplest example of the situation where the target POVM elements are linearly dependent
is the qubit measurement in the BB84 QKD protocol.
There, the sum of the elements of either basis is proportional to the identity operator and therefore
it is not hard to find scalars $\alpha_i$ such that Eq.~(\ref{eq:targetlindep}) holds.

Using the vectorization of the POVM elements it is convenient to rewrite Eq.~(\ref{eq:lineardependancegeneral})
as
\be
\sum_{i}\alpha_i \vek{F_{M}^{(i)}}=0
\Leftrightarrow\sum_{i,j}\alpha_{i}\PP_{ij}\vek{F_{B}^{(j)}}=0,
\ee
where the vectorization $\vek{}$ gives an isomorphism between the linear bounded operators
and vectors in corresponding spaces.
Considering $\vek{F_{B}^{(i)}}$ for any $i$ as an $i$-th column of a matrix
and writing $\vec{\alpha}=(\alpha_1,\dots,\alpha_{N_Q})^T$ gives:
$
\FF_{Q}\vec{\alpha}=0\Leftrightarrow \FF_{M}\vec{\alpha}=0
\Leftrightarrow\FF_{B}\left(\PP^T\vec{\alpha}\right)=0.
$
In summary, we have the following observation.
\begin{obs}(Valid CPP schemes)
A valid postprocessing that allows for the existence of a squashing map is a stochastic matrix such that its transpose maps
the null space of the matrix, built from the vectorizations of the basic POVM elements $\FF^{(i)}_{B}$,
onto the null space of the matrix, built from the vectorizations of the target POVM elements $\FF^{(i)}_{Q}$:
\be
\PP^T: \nullsp{\FF_{Q}} \rightarrow \nullsp{\FF_{B}}.
\ee
\end{obs}

Note that this condition incorporates both the validity of a CPP scheme, as stated in Definition \ref{CPPdefn}, and
the requirement that the linear dependencies of the POVM elements on the truncated and initial Hilbert spaces has
to be respected. Hence if this condition is satisfied then there always exists a linear map connecting the full and
the target measurements.

\subsection{Determining the existence of a squashing map: complete positivity}
The last constituent of a squashing model is the positivity of the squashing map (the linear map from the previous section).
In order to check for positivity, we employ the Choi-Jamio{\l}kowski
isomorphism \cite{jamiolkowski72a,choi75a}
\be
\tau\equiv\eins_Q\otimes\Lambda^{\dagger}_{B}\left(\proj{\psi^+}\right)\geq 0,
\label{CJmatrix}
\ee
where $\tau$ is called the Choi matrix and $\ket{\psi^+}=\frac{1}{\sqrt{d_Q}}\sum_i\ket{i}\ket{i}$ is a normalized maximally entangled state. This isomorphism is formulated directly from the action of the adjoint map $\Lambda^\dagger_{B}$.
The property of complete positivity of the map $\Lambda^\dagger_{B}$ is then equivalently expressed
by the positivity of the corresponding Choi matrix.

The linear constraints on the map $\Lambda^\dagger_{B}$ are best expressed in terms of the so-called natural
representation $\tau^R$ (see Appendix A for details).
The direct link between the full and the target measurement in the natural representation is given by
\be
\tau^R \vek{F_Q^{(i)}}=\vek{F_M^{(i)}}.
\ee
Note that this automatically includes the condition for the map $\Lambda^\dagger_{B}$ to be unital, as can be checked
by summing over the index $i$.

As a consequence, one can reformulate the problem of looking for
a completely positive squashing map as a special instance of
semi-definite programming (feasibility problem) \cite{vandenberghe96a}:

\bea
\textnormal{Find }\tau &\geq 0,\nonumber\\
\textnormal{s.t. }\tau^R &\vek{F_Q^{(i)}}=\vek{F_M^{(i)}}\label{sdpsquash}.
\eea

It is clear from the provided construction that the positivity of the Choi matrix (and therefore
of the squashing map) crucially depends on the choice of the classical postprocessing. In fact,
as we will see in the next section and in Section \ref{ssection:6statealternativeCPP},
if positivity is not achieved by the basic measurement it can be always repaired by
choosing another valid postprocessing, though this will typically be at some price in
terms of protocol performance.

\subsection{Enforcing existence of squashing maps by noisy post-processing}\label{noisyCPPs}
In this section we point out two important facts: {\it (i)} We can always find a squashing model for any pair of
target and basic measurements by choosing a suitable (although very noisy) post-processing which we call a {\it trivial squashing
model} (Proposition \ref{alwayssquash}). {\it (ii)} Despite the fact that
this trivial squashing model might, at first sight, appear useless, we can use its completely positive squashing map
in order to restore the positivity of another squashing map that appears to be non-positive and therefore construct
a {\it non-trivial squashing model}. That will be the essence of the {\it restoring theorem} (Theorem \ref{mixedsquashmap}).

To be more specific we provide an example for how a noisy squashing model can be used
and then turn to the general case.
A typical situation where we look for squashing models has the property that
the target measurement corresponds to a restriction of the
basic measurement to some simple subspaces, for example those of single-photon signals. In these cases,
one will usually try to make a smart choice of postprocessing, namely such that the
postprocessing retains this property, i.e.~the restriction of the full measurement to the
specified subspaces results in the target measurement.
An example of such a CPP scheme arises in the context of the six-state measurements (defined in Section \ref{ssection:6stateact})
where one makes a random assignment of double-clicks (when two detectors fire simultaneously),
while keeping the single click events unchanged.
As we will see, the squashing map constructed for this CPP scheme is not completely positive (Section \ref{sssection:6stateNCP}). However, this positivity problem can be overcome by statistically mixing
the smart postprocessing (where single clicks events are unchanged) with a noisy
postprocessing that will also reassign single click outcomes (see Section \ref{ssection:6statealternativeCPP}).

To start out, we introduce a postprocessing which allows the existence of a trivial squashing model.

\begin{figure}
\includegraphics[width=.49\columnwidth]{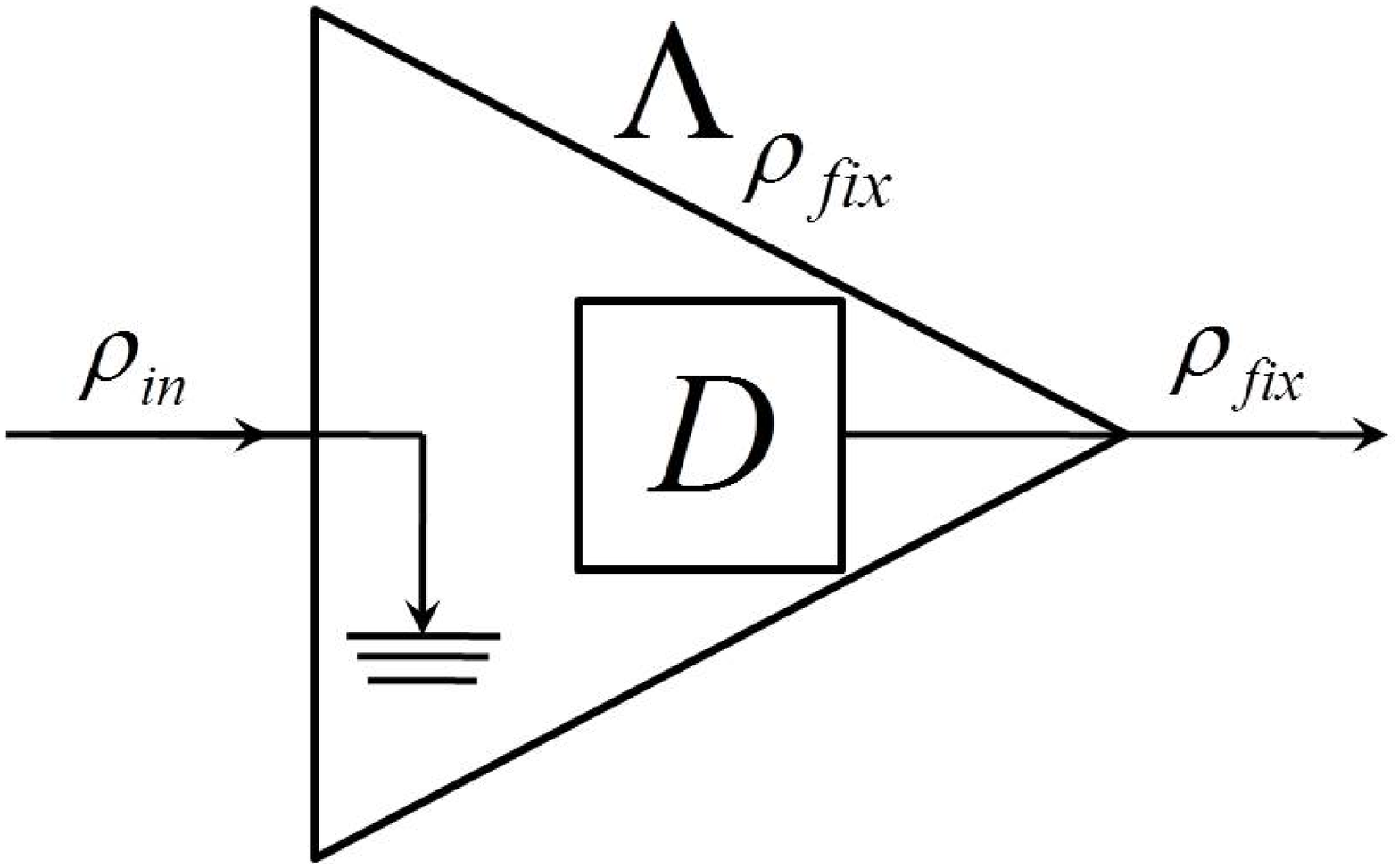}
\caption{Squashing map $\Lambda_{\vr_{fix}}$ that disregards the input and outputs a fixed
state $\vr_{fix}$, which is prepared by a device $D$. \label{unisquash}}
\end{figure}

\begin{prop}\label{alwayssquash}(Trivial squashing map)
Let $F_B^{(i)}$ be any complete set of POVM elements that characterize the basic measurement and let $F_Q^{(j)}$ be
some complete set of POVM elements that characterize the target measurement.
Let the classical postprocessing be such that it redistributes all basic events according to some a priori
fixed probabilities, which are derived from a density matrix $\vr_{fix}$ as
$p_Q^{(i)}=\tr{\vr_{fix}F_Q^{(i)}}$ and which do not depend on the input state. Then there
always exists a squashing model such that its map $\Lambda_{\vr_{fix}}$ acts trivially on any input state $\vr_{in}$,
i.e.~$\Lambda_{\vr_{fix}}[\vr_{in}]=\vr_{fix}$.
\end{prop}
{\it Proof:} The statement of the proposition is a link between positivity of a squashing map for any type
of basic and target measurements and a certain classical postprocessing.
An idea of how such a post-processing scheme can be constructed, and which
squashing map it corresponds to, is presented in Fig.~\ref{unisquash}.
We apply a postprocessing which ignores the
measurement result and assigns an outcome with fixed a priori probabilities
$p_Q^{(i)}=\tr{\vr_{fix}F_Q^{(i)}}$ compatible with some fixed quantum state $\vr_{fix}$.
We define a map $\Lambda_{\vr_{fix}}$ such that
$\Lambda_{\vr_{fix}}[\vr_{in}]=\vr_{fix}$ for any $\vr_{in}$. By construction this map is
completely positive and fulfills the linear constraints of Eq.~(\ref{linconstOperForm}).
\qed

Before we show how this type of the squashing map is useful we need to point out an important
property of its Choi matrix.
\begin{rem}(Properties of the Choi matrix for the trivial squashing map)\label{noisyfullrank}
The minimum eigenvalue of the Choi matrix of the trivial squashing map $\Lambda_{\vr_{fix}}$ is
proportional to the minimum eigenvalue of the state $\vr_{fix}$ with
the coefficient of proportionality $1/d_Q$, where $d_Q$ is the dimension of the target
Hilbert space $\HH_Q$.
\end{rem}
{\it Proof:}
As defined in Proposition \ref{alwayssquash} $\Lambda_{\vr_{fix}}[\vr]=\vr_{fix}$ for any $\vr$.
Therefore the adjoint map $\Lambda^{\dagger}_{\vr_{fix}}$ must satisfy
\be
\tr{\vr_{fix}\OO}=\tr{\vr\Lambda^{\dagger}_{\vr_{fix}}[\OO]}
\ee
for any bounded operator $\OO$. This implies that
\be
\Lambda^{\dagger}_{\vr_{fix}}[\OO]=\tr{\vr_{fix}\OO}\eins_M
\ee
which is reminiscent of the completely depolarizing map.
Then the Choi matrix of the adjoint map is explicitly given by
\bea
&\tau_{\vr_{fix}}=\eins_Q\otimes\Lambda^{\dagger}_{\vr_{fix}}(\proj{\psi^+})\nonumber\\
&=\frac{1}{d_Q}\sum_{ij}\ket{i}\bra{j}\otimes\Lambda^{\dagger}_{\vr_{fix}}[\ket{i}\bra{j}]\label{eq:taunoise}\\
&=\frac{1}{d_Q}\sum_{ij}\bra{j}\vr_{fix}\ket{i}\ket{i}\bra{j}\otimes\eins_M=\frac{1}{d_Q}\vr^T_{fix}\otimes\eins_M,\nonumber
\eea
and so the assertion follows.
\qed

It is clear that choosing such a postprocessing and constructing such a squashing model
is not very clever, because one looses all useful data
from the performed measurement and eventually winds up with a squashing model
that produces only noise.

Nevertheless this tool turns out to be very useful in particular cases. In fact,
as we will see now, Proposition \ref{alwayssquash} and Remark \ref{noisyfullrank}
together imply that a positive squashing map can always be found by
introducing some amount of noise on the measurement data.

\begin{thm}(Restoring theorem)\label{mixedsquashmap}
Let $F_B^{(i)}$ and $F_Q^{(i)}$ be the basic and the target POVM elements of the
corresponding measurement devices respectively.
Let $\tau$ be a Choi matrix, such that for classical
postprocessing $\PP$ the linear constraints in Eq.~(\ref{sdpsquash}) are satisfied, but $\tau\ngeq 0$.
Then there exists a state $\vr_{fix}$ and another postprocessing
$\PP'$ with intermediate amount of the added noise $p$, which provides
a squashing map with the Choi matrix
\be
\tau'(p) = (1-p)\tau + p\tau_{\vr_{fix}}
\label{mixedChoi}
\ee
that is positive semi-definite whenever $p$ and $\vr_{fix}$ are chosen according to
\be
\lambda_{\textnormal{min}}(\tau')\geq (1-p)\lambda_{\textnormal{min}}(\tau) + p\lambda_{\textnormal{min}}(\vr_{fix})/d_Q\geq 0.
\label{eq:eigvalineqnoise}
\ee
\end{thm}
{\it Proof:} First we note that due to Proposition \ref{alwayssquash} a completely noisy postprocessing
exists that allows for a completely positive squashing map.
Now we look for an intermediate postprocessing that introduces less noise and where the
corresponding Choi matrix is still positive.

This intermediate postprocessing will be chosen as a probabilistic mixture of the
postprocessing $\PP$ and the noisy postprocessing
$\PP_{\textnormal{noise}}$ from Proposition \ref{alwayssquash},
\be
\PP' = (1-p)\PP + p\PP_{\textnormal{noise}}.\label{noisyscheme}
\ee
For the full measurement POVM elements this implies
\be
F_M^{\prime(i)} = (1-p)F_M^{(i)} + p F_{\textnormal{noise},M}^{(i)},
\ee
so that we can construct an adjoint of the squashing map
\be
\Lambda^{\prime\dagger}[F_Q^{(i)}] =(1-p)\Lambda^{\dagger}[F_Q^{(i)}] + p\Lambda^{\dagger}_{\vr_{fix}}[F_Q^{(i)}].
\ee

The choice of the $\vr_{fix}$ is of crucial importance here. We choose $\vr_{fix}$ to have a full rank,
so that the eigenvalues of the Choi matrix in Eq.~(\ref{eq:taunoise}) are all strictly positive.

For $p=1$ the new squashing map $\Lambda'$ is completely positive (Proposition \ref{alwayssquash}).
For the rest of the parameter values we investigate the positivity of
$\tau'=\eins_Q\otimes\Lambda^{'\dagger}(\proj{\psi^+})$, which satisfies
\be
\tau'(p) = (1-p)\tau + p\tau_{\vr_{fix}}
\nonumber
\ee
and where $\tau_{\vr_{fix}}$ is given explicitly by Eq.~(\ref{eq:taunoise}).
Since the second term in the last
equation is strictly positive, which is guaranteed by our choice of $\vr_{fix}$ and Remark \ref{noisyfullrank},
there exists a value of $p$ that is smaller than the trivial value $p=1$
and for which we still find  $\tau' \geq 0$ and therefore a nontrivial
complete positive squashing map $\Lambda'$.

The amount of noise $p$ that guarantees the positivity of $\tau'$ can be
determined by comparing the minimal eigenvalue of $\tau$ (which is negative) and the
minimal eigenvalue of $\vr_{fix}$ (which is positive).
It follows from one of Weyl's inequalities (see for example Chapter III.2 in \cite{bhatia97a})
that the minimum eigenvalue of the sum of Hermitian matrices is lower bounded by the sum of the minimal eigenvalues
of each term which implies the positivity condition in Eq.~(\ref{eq:eigvalineqnoise})
\be
\lambda_{\textnormal{min}}(\tau')\geq (1-p)\lambda_{\textnormal{min}}(\tau) + p\lambda_{\textnormal{min}}(\vr_{fix})/d_Q\geq 0.
\nonumber
\ee
\qed

To close the section we comment on the non-triviality of the map constructed in the last theorem.
The existence of a positive $\tau'(p)$ for some $p<1$ implies that there exists a CPP scheme
that allows a physical map which preserves some quantum properties of the input state (a specific example will
be given in Section \ref{ssection:6statealternativeCPP}).
This is especially important to know for applications such as verification of
entanglement, which has, for example, applications as necessary conditions for QKD \cite{curty04a}.

\section{Reductions for squashing models for linear optical devices with threshold detectors}\label{linoptimpl}
The description of a general linear optical measurement device, that is, a device in which input modes
undergo a linear transformation before entering detectors, can be rather
complicated. A full description should include several different degrees of freedom. For example, incoming light can consist of several spatially separated or overlapping wave packets with an
arbitrary number of photons, each with its various polarization or frequency.
However, we restrict our analysis to measurement devices that only respond to particular degrees of freedom.
This means that they are invariant in their statistics with a change in degrees of freedom they do not measure.
Specifically, in what follows, we consider measurement devices that can have different statistics given a change in
photon number and polarization (Sections \ref{ssection:BB84active}-\ref{rubenswork})
or in time (see Section \ref{sectm}) or in photon number and relative phase between two spatial modes
(Section \ref{Application}). This implies that these measurement devices are invariant under changes in
all other degrees of freedom, such as frequency.

The existence of common attributes in optical measurement devices makes the
application of the general framework discussed in the previous sections easier.
The fact that one usually uses threshold detectors turns out to be especially helpful. As we will
see shortly, this allows to decompose the Hilbert space of the incoming signal and to construct the
squashing map for $N$-photon input states for each $N=0,1,2,3,...$ independently.
This will be the first reduction for the linear optical measurement devices.

In addition, we show that for a special type of basic and target measurements there exists a particular
CPP scheme that allows further decomposition in each of the $N$-photon subspaces that
substantially simplifies the squashing model analysis.

\subsection{Quantum non-demolition (QND) measurements and $N$-photon subspaces}\label{section:QND}
One essential trick in managing the analysis of squashing models connecting infinite-dimensional
mode spaces to finite-dimensional target measurements consists of exploiting the fact that
the basic POVM elements of linear optical measurement devices with threshold detectors commute with
the POVM elements of the QND measurement of the total number of photons.
This allows the reduction of the problem of analyzing input states on an infinite-dimensional Hilbert space
to the problem of analyzing input states on an infinite number ($N=0,1,2,3,...$) of finite-dimensional Hilbert spaces
independently. Formally, we can write
\be
\textnormal{QND}:\vr\mapsto\bigoplus_{N=0}^{\infty}\vr_{N}.
\label{eq:QNDblockdiagstate}
\ee

Based on what was laid out above, we can assume without loss of generality that
the squashing map first performs a QND measurement of the total photon number,
thus turning the input state into a block-diagonal form with respect to the photon number subspaces.
It implies that we can check for the existence of a squashing model for each subspace
separately. An important note here is that a CPP scheme has to be fixed before we start to search
for a squashing model on the infinite family of finite-dimensional Hilbert spaces. Each of these
squashing models have to share a common CPP scheme.

All of the above can be summarized as follows.
\begin{obs}(Reduction 1: QND)\label{obs:QNDreduction}
For a linear optical measurement device with threshold detectors, any
squashing map has a block-diagonal form with respect to the photon number subspaces:
\be
\Lambda[\vr]\stackrel{\textnormal{QND}}{=}
\Lambda\left[\bigoplus_{N=0}^{\infty}\vr_N\right]=\bigoplus_{N=0}^{\infty}\Lambda_N[\vr_N].
\label{eq:QNDreduction}
\ee
\end{obs}

Note that for any $N$ the map $\Lambda_N$ is characterized by the same target measurement and
its adjoint maps the target POVM elements onto the full measurement POVM elements projected onto the
$N$-photon subspace.

One immediate consequence of the QND measurement is the fact that one can split off the vacuum component
and only consider states with $N\geq1$. Indeed, for $N=0$, one can always choose
$\Lambda_0[\vr_0]=\ketbraaa{\textnormal{vac}}{Q}{}$, where $\ket{\textnormal{vac}}$
is the vacuum state in the target Hilbert space.
Therefore the squashing map will output a vacuum state
whenever the outcome of the QND measurement is zero and we can restrict ourselves to the case where $N\neq 0$:
\be
\bigoplus_{N=0}^{\infty}\Lambda_N[\vr_N]=\ketbraaa{\textnormal{vac}}{Q}{}\oplus\bigoplus_{N=1}^{\infty}\Lambda_N[\vr_N].
\ee
This will be referred to as the vacuum flag structure of the squashing map.
Note that the map $\Lambda_0$ is applied if and only if the outcome of the QND measurement is 0. Later on in Section
\ref{ssection:BB846statepassive} we will become acquainted with another map, which outputs a vacuum state on
the target Hilbert space, no matter what the input is. This ``vacuum map'' should not be confused with
the vacuum flag, whose sole role is to split off the vacuum component of the signal.

\subsection{Reduction for natural CPP schemes}\label{section:OrthoSubspace}
For measurement devices with threshold photodetectors,
for which the target measurement can be described as a restriction of the
basic measurement to the single-photon subspace,
the smart choice of a CPP scheme is such that the scheme does not
affect the events which could have come from single-photon signals.
These events are single-clicks. Separating the single-clicks from the rest of events in this
way will lead to a CPP scheme which assigns all multi-clicks to some single-clicks without
performing any operation on single-click outcomes,
so that overall full measurement POVM elements are of the form
\be
F_M^{(i)}=F_{B,\textnormal{single}}^{(i)}+\sum_j \PP_{ij}F_{B,\textnormal{rest}}^{(j)}.
\label{eq:PPorthoCPP}
\ee

\begin{figure}
\includegraphics[width=.99\columnwidth]{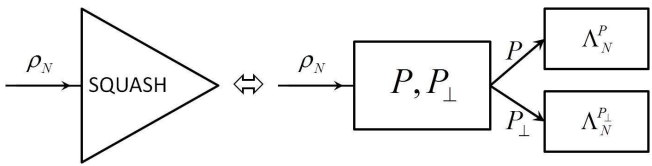}
\caption{Action of the squashing map for the special type of CPP scheme preserving the
single-click events. The squashing map can be modeled
as a photon number measurement followed by a projective
measurement onto a subspace spanned by the pure states that can trigger only single-click events.
The number of such states does not depend
on the photon number $N\geq1$. Depending on
the outcome of these measurements, one either proceeds with
a low-dimensional squashing operation $\Lambda_N^P$
or outputs a completely mixed qubit state. \label{fig:PPortho}}
\end{figure}

For generic linear optical devices, $F_{B,\textnormal{single}}^{(i)}$ has a form of a rank-1 projector on
some state $\ket{\Psi_{B,\textnormal{single}}^{(i)}}$. If we denote a space spanned by
single-click states by $P=\textnormal{span}\{\ket{\Psi_{B,\textnormal{single}}^{(i)}}\}$ then any
state from its orthogonal compliment $P_{\bot}$ triggers a multi-click with certainty.
If the projection on $P$ commutes with full measurement POVM elements
we can investigate the existence of the squashing map for $P$ and $P_{\bot}$ separately.
Schematically this situation is represented in Fig.~\ref{fig:PPortho}.

In summary we have the following observation.
\begin{obs}(Reduction 2: Single-click subspace)\label{obs:PPorth}
If a linear optical device with threshold photodetectors is such that
\bi
\item[i)] The reduction of the basic POVM elements to a single
photon subspace provides target measurement POVM elements
\item[ii)] The projection on the space $P$, spanned by the states that can trigger only
single click events, commutes with the full measurement POVM elements,
\ei
then there exists a CPP scheme that allows a decomposition of the
squashing map of the form
\be
\Lambda_N = \Lambda_{P,N} + \Lambda_{P_{\bot},N}.
\label{eq:PPorthodecomp}
\ee
\end{obs}
Both $\Lambda_{P,N}$ and $\Lambda_{P_{\bot},N}$ map $N$-photon states to states
on the same target Hilbert space and fulfil the same set of linear constraints.

\section{Squashing model for a measurement device used in the BB84 QKD protocol}\label{ssection:BB84active}
In this section we consider a measurement device which is used in the optical implementation of the most prominent
QKD protocol: the BB84 protocol \cite{bennett84a}.
This device has been introduced in \cite{beaudry08a} as a standard example used to introduce squashing models.

We start off by providing a short background on the measurement device in the BB84 QKD protocol,
where an observer actively makes the choice of the measurement basis.

\subsection{Active detection scheme for the BB84 measurement}
\begin{figure}
\includegraphics[width=.99\columnwidth]{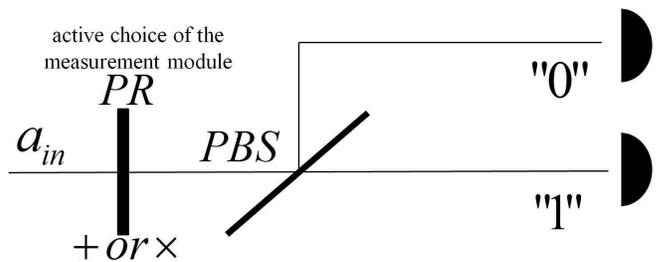}
\caption{Active detection scheme. The observer possesses two detector modules and a polarization rotator, which
is used to actively choose one or the other detector module.
The detector modules are made up of polarizing beam splitters that are able to
discriminate two orthogonal linearly polarized
modes and ideal threshold detectors for each mode. \label{PBSBB84}}
\end{figure}

In the active detection scheme for the BB84 measurement the observer has two detector modules, each adjusted to one
of the polarization bases $\alpha$ (see Fig.~\ref{PBSBB84}). Before the measurement is performed one has to decide
which detector module will be used. This represents the active nature of the detection scheme.

Note that there is no notion of an a priori probability distribution that governs the choice of the measurement
basis yet, i.e.~the observer has no classical ``coin" at his disposal and therefore no randomness for the basis choice.
A measurement with such randomness will be discussed in Section \ref{ssection:BB84biasedactive}.

Each detector module is a polarization analyzer and consists of two detectors monitoring two outputs
of a polarizing beam splitter, which is able to discriminate between two orthogonal polarizations
of linearly polarized light.

In every measurement the observer will register four different events: no click (vac), single-click (sc) in one of the detectors,
and two different double-clicks (dc), when both detectors ($"0"$ and $"1"$) fire.
Assuming ideal threshold detectors, no photon losses and
no dark counts the observed events are described by the following POVM elements
\bea
F_{\textnormal{vac}}=&\sum_{\alpha=+,\times}\ketbraa{0,0}\nonumber\\
F^{i,\alpha}_{\textnormal{sc}}=&\sum_{N=1}^{\infty}\ketbraaa{N}{i,\alpha}{},\label{sceffects0}\\
F^{\alpha}_{\textnormal{dc}}=
&\sum_{N_0,N_1=1}^{\infty}\ketbraaa{N_0,N_1}{\alpha}{},\nonumber
\eea
where $\ket{N}_{i,\alpha}$ denotes a state with $N$ photons in the mode $i$ of the $\alpha$-polarized
{\it incoming} light (cf.~\cite{nl99a}).

Note that for each choice of $\alpha$ one has a {\it complete set} of POVM elements and {\it no} classical
probability that may describe the basis choice.

\subsection{BB84 measurement: reduction of the squashing model}\label{ssection:activeBB84reduction}
We start off by defining target POVM elements. These correspond to a measurement
on zero and single photon Hilbert spaces and are given by
(cf.~Eq.~(\ref{sceffects0}))
\bea
F_{\textnormal{vac}}=&\sum_{\alpha=+,\times}\ketbraa{0,0}\nonumber\\
F^{(0,\alpha)}_{1} &=\ket{1,0}_{\alpha}\bra{1,0},\label{BB84qutrit}\\
F^{(1,\alpha)}_{1} &=\ket{0,1}_{\alpha}\bra{0,1},\nonumber
\eea
where $\alpha\in \{+,\times\}$ is a label for the basis choice of the
polarizing beam splitter.

For a general input state we apply reductions from Section \ref{linoptimpl} in order to reduce the problem.
First, in virtue of what is discussed in Section \ref{section:QND}, we consider an input
state of the BB84 measurement device that contains of $N$ photons and by using the flag structure
of the vacuum events, we split off the vacuum component. This also simplifies the target space and makes
it a space of a qubit with the following POVM elements
\be
F^{(0,\alpha)}_{1} =\ket{1,0}_{\alpha}\bra{1,0},\;
F^{(1,\alpha)}_{1} =\ket{0,1}_{\alpha}\bra{0,1}.\label{BB84target}
\ee

Second, the target POVM elements (Eq.~(\ref{BB84target})) are restrictions of the general
basic POVM elements to the single-photon subspace.
Therefore we can apply the results of Section \ref{section:OrthoSubspace} and choose a CPP scheme
that does not affect the single-clicks in order to be able to decompose $\Lambda_N$ into $\Lambda_{P,N}$
and $\Lambda_{P_{\bot},N}$.

To perform this decomposition we fix the CPP scheme by randomly (with equal probability) assigning each of the double-click
events to a single-click event within the same basis.
It follows directly from Eq.~(\ref{sceffects0}) that the full measurement POVM elements
on the $N$-photon subspace are
\bea
F^{(b,\alpha)}_{N} &= F^{(b,\alpha)}_{\textnormal{sc},N} + \halbe F_{\textnormal{dc},N}^{\alpha}\nonumber\\
&=\frac{(-1)^b}{2} \left( \ket{N,0}_{\alpha}\bra{N,0} - \ket{0,N}_{\alpha}\bra{0,N}\right)
+ \frac{\eins_N}{2},
\label{BB84povms}
\eea
where $b\in \{0, 1\}$ corresponds to the 0 or 1 outcome of the detection module,
and $\ket{l, k}_{\alpha}$ is a two mode
Fock state with photon numbers $l$ and $k$ with respect
to the polarization mode basis $\alpha$. It is straightforward to see that the restriction of these elements
to the $N=1$ subspace exactly reproduces the target POVM elements.

Now we see that the full measurement POVM elements Eq.~(\ref{BB84povms}) have the same structure
of the POVM elements in Eq.~(\ref{eq:PPorthoCPP}). Moreover, projections on the spaces
$P=\textnormal{span}\{\ket{N, 0}_{\alpha}, \ket{0,N}_{\alpha}\}_{\alpha=+,\times}$ and $P_{\bot}$
commute with the full measurement POVM elements and therefore we can apply Observation
\ref{obs:PPorth} in order to search for a squashing map in $P$ and $P_{\bot}$ separately.

\subsection{BB84 measurement: positivity of the squashing map}

For any $N\geq1$ we can choose $\Lambda_{P_{\bot},N}$ to be a trivial map $\Lambda_{\varrho_{fix}}$ from
Proposition \ref{alwayssquash} and the choice $\varrho_{fix}=\eins_Q/2$. This is in accordance with the chosen
CPP scheme, i.e.~the probability condition in Eq.~(\ref{probcond}) is fulfilled.
Thus we have determined the squashing map on the space $P_{\bot}$ and all that is left to find is a squashing map for
the subspace $P$, whose dimension in this case does not exceed 4 (it is 2 for $N=1$, 3 for $N=2$ and 4 for $N\geq 3$).

As we mentioned in Section \ref{genstrat} we need two ingredients for this: {\it (i)} we need
to construct a linear map preserving linear dependencies as in Eq.~(\ref{eq:lineardependancegeneral}) and
{\it (ii)} we need this map to be completely positive.

We start off by writing down the linear constraints, which are respected by
the chosen CPP scheme
\be
\Lambda_{P,N}^{\dagger}\left[F^{(b,\alpha)}_{1}\right]=F^{(b,\alpha)}_{P,N},\: \alpha\in\{+,\times\},
\label{linconstrBB84}
\ee
where $F^{(b,\alpha)}_{1}$ are the target measurement POVM elements as in Eq.~(\ref{BB84target})
and $F^{(b,\alpha)}_{P,N}$ are the full measurement POVM elements restricted to the subspace $P$.

Using Eq.~(\ref{BB84target}) and Eq.~(\ref{BB84povms}) it is not hard to see that the
linear dependencies are satisfied
\be
F^{(0,\alpha)}_{1}+F^{(1,\alpha)}_{1}=\eins_Q
\Leftrightarrow
F^{(0,\alpha)}_{P,N}+F^{(1,\alpha)}_{P,N}=\eins_{P,N},\forall\alpha.
\ee
Therefore there exists a linear map $\Lambda_{P,N}^{\dagger}$ as in Eq.~(\ref{linconstrBB84}).

The complete positivity of $\Lambda_{P,N}^{\dagger}$ is proven by directly
checking the non-negativity of the Choi matrix. First, we
use the decomposition of the maximally entangled state
in terms of Pauli matrices
\be
\ketbra{\psi^+}=\frac{1}{4}\left(\eins_Q\otimes\eins_Q + \sum_{\alpha = x,y,z}\sigma^T_{\alpha}\otimes\sigma_{\alpha}\right).
\label{eq:mesinpaulis}
\ee
Therefore the Choi matrix is given explicitly by
\begin{widetext}
\be
\tau_{P,N} =\eins\otimes\Lambda_{P,N}^{\dagger}\left(\ketbra{\psi^+}\right)=\frac{1}{4}\left(\eins_Q\otimes\eins_M + \sum_{\alpha = x,y,z}\sigma^T_{\alpha}\otimes\Lambda_{P,N}^{\dagger}(\sigma_{\alpha})\right).
\ee
\end{widetext}
Second, we note that the Pauli matrices $\sigma_x$ and $\sigma_z$ can be written in terms of the target POVM elements: $\sigma_{\alpha}=F^{(0,\alpha)}_{1}-F^{(1,\alpha)}_{1}$, $\alpha=x,z$. The action of the
$\Lambda_{P,N}^{\dagger}$ on $\sigma_y$, however, is not fixed by our linear constraints
and we can use this freedom in order to enforce the positivity of the Choi matrix $\tau_{P,N}$.

For the upcoming discussion it is convenient to decompose $\tau_{P,N} = \tau_{P,N,\textnormal{fix}}+\tau_{P,N,\textnormal{open}}$, with
\be
\tau_{P,N,\textnormal{open}}=\sigma_y^T\otimes\Lambda_{P,N}^{\dagger}\left(\sigma_y\right).
\ee
In order to check the positivity of $\tau_{P,N}$ we will consider its matrix representation $M(\tau_{P,N})$
using the non-orthogonal basis vectors
\be
\left\{\ket{\psi_i}\otimes\ket{j}\right\}_{j=0,1},\: \ket{\psi_i}\in\{\ket{N,0}_{\alpha},\ket{0,N}_{\alpha}\}_{\alpha=+,\times}.
\label{BB84basis}
\ee

The matrix $M(\tau_{P,N,\textnormal{fix}})$ only has real entries
(its explicit form is given by Eq.~(\ref{BB84exact}) in the appendix).
The properties of the matrix $M(\tau_{P,N,\textnormal{open}})=\sigma_y^T\otimes M\left(\Lambda_{P,N}^{\dagger}(\sigma_y)\right)$
can be specified further: first, without loss of generality, we
can assume that $M(\tau_{P,N,\textnormal{open}})$ only has real entries. If there
exists a complex solution $M(\tau_{P,N,\textnormal{open}})$ for $M(\tau_{P,N,\textnormal{fix}})
+M(\tau_{P,N,\textnormal{open}})\geq 0$
then its complex conjugate is also a solution. Then by linearity the equal weighted average is also a solution and it is a real matrix.
Second, since the open part $\tau_{P,N,\textnormal{open}}$ is Hermitian (otherwise the Choi matrix would
have complex eigenvalues), we can write
$M\left(\Lambda_{P,N}^{\dagger}(\sigma_y)\right)=iS$ where $S$ is some skew symmetric matrix
with 6 real entries as free parameters. These free parameters can be found such that $M(\tau_{P,N})\geq 0$
holds (see Eq.~(\ref{BB84solutions})). Therefore, there exists a positive $\tau_{P,N}$ which maps the specified
target measurements to the corresponding full measurements.

This implies that there is a squashing map on the space $P$, which is completely positive and
fulfills the linear constraints in Eq.~(\ref{linconstrBB84}). As pointed out above, the squashing
map on the complementary space $P_{\bot}$ also exists. Therefore, for the choice of the classical postprocessing
we made, we provided a squashing map for the target measurement in Eq.~(\ref{BB84target}).
This squashing map has been found also by Tsurumaru and
Tamaki \cite{tsurumaru08a}.

In summary in this section we proved the following.
\begin{thm}(Squashing model for BB84 measurement with active basis choice)
There exists a squashing model with the qubit target measurement for the BB84 measurement
with active basis choice and random
equiprobable assignment of double-clicks to one of the outcomes
(i.e.~independent from how often one decides to choose one or the other basis).
\end{thm}

\section{Active detection scheme for a six-state measurement}\label{ssection:6stateact}
In this section we will focus on a squashing model for a measurement device
which is used in optical implementations of the six-state QKD protocol \cite{bruss98a}.

The six-state measurement is similar to the BB84
measurement, except that there is a third setting to the polarizing
beam splitter which splits photons according to
a circular basis (labeled as $y$). The measurement basis is chosen actively by the
observer.

\subsection{Six-state measurement: reduction of the squashing model}
In full analogy to the BB84 measurement we want to make use of the
reductions in Section \ref{linoptimpl}. First of all we reduce
the problem and consider $N$ photon incoming signals for $N=1,2,...$.

For the six-state measurement device we choose the target measurement to
be a measurement on the zero and single photon Hilbert space. After splitting off
the vacuum component, the target POVM elements are
\be
F^{(0,\alpha)}_{1} =\ket{1,0}_{\alpha}\bra{1,0},\;
F^{(1,\alpha)}_{1} =\ket{0,1}_{\alpha}\bra{0,1},
\label{sixstatetarget}
\ee
with $\alpha \in \{x, y, z\}$.

As with the BB84 measurement these POVM elements are restrictions of the basic POVM
elements, which suggests to choose the same type of classical postprocessing
as we did for the BB84 protocol: the postprocessing of double click events
is randomly assigned again to either single detection events. The probabilities
of the ``0" and ``1" assignments are equal, $p=1/2$. This CPP scheme fixes the full measurement
POVM elements to
\be
F^{(b,\alpha)}_{N}=\frac{(-1)^b}{2} \left( \ket{N,0}_{\alpha}\bra{N,0} - \ket{0,N}_{\alpha}\bra{0,N}\right)
+ \frac{\eins_N}{2},
\label{eq:6statefull}
\ee

This CPP scheme allows us to apply the second reduction and restrict our
search to a six-dimensional subspace $P$ spanned by
$\{\ket{N, 0}_{\alpha}, \ket{0,N}_{\alpha}\}$, $\alpha=x,y,z$, and its complement $P_{\bot}$.
Similarly to the active BB84 measurement the projections onto $P$ and $P_{\bot}$ commute
with the full measurement POVM elements in Eq. (\ref{eq:6statefull}).

\subsection{Six-state measurement: squashing map}\label{sssection:6stateNCP}
The linear constraints on the squashing map in
the six-dimensional subspace $P$ are given by:
\be
\Lambda_{P,N}^{\dagger}\left[F^{(b,\alpha)}_{1}\right]=F^{(b,\alpha)}_{P,N},\: \alpha\in\{x,y,z\}.
\label{linconstr6state}
\ee

In this case, one can follow the calculation for the BB84 protocol.
The only difference is that the matrix $\tau_{P,N}$, that represents the squashing
map, is completely determined by the linear constraints
since the measurement operators $F^{(b,\alpha)}_{1}$ form a complete basis
for their Hilbert space.
However, it can be easily seen that $\Lambda_{P,N}$
cannot be positive. First, we can
write the adjoint squashing map $\tau_{P,N}=\eins\otimes\Lambda_{P,N}^{\dagger}\left(\ketbra{\psi^+}\right)$
as before. Since the qubit measurements of the six-state
protocol are complete, we can write
\begin{widetext}
\be
\tau_{P,N}=\eins\otimes\Lambda_{P,N}^{\dagger}\left(\ketbra{\psi^+}\right)=\frac{1}{4}\left(\eins_Q\otimes\eins_{P,N} + \sum_{\alpha=x,y,z} \sigma^T_{\alpha}\otimes(F^{(0,\alpha)}_{P,N}-F^{(1,\alpha)}_{P,N})\right).
\label{ChoiMatrix6state}
\ee
\end{widetext}
As in the BB84 case, we can directly apply $\Lambda_{P,N}^{\dagger}$ to the
second subsystem that does the map $F^{(b,\alpha)}_{1}\mapsto F^{(b,\alpha)}_{P,N}$. In this case, the Choi matrix
$\tau_{P,N}$ is completely fixed and has no free parameters since the linear constraints in Eq.~(\ref{linconstr6state})
have to be respected.

However, the matrix $\tau_{P,N}$ has negative eigenvalues. By writing $\tau_{P,N}$ in a basis
of non-orthogonal vectors, which is analogous to that for the BB84 measurement device in Eq.~(\ref{BB84basis}):
\be
\left\{\ket{\psi_i}\otimes\ket{j}\right\}_{j=0,1},\: \ket{\psi_i}\in\{\ket{N,0}_{\alpha},\ket{0,N}_{\alpha}\}_{\alpha=x,y,z},
\label{6statebasis}
\ee
we can calculate the minimum eigenvalues directly. For example in the three photon subspace, the state
\be
\ket{\theta_-}=\frac{1}{\sqrt{2}}\left(\ket{3,0}_z\otimes\ket{1}-\ket{0,3}_z\otimes\ket{0}\right)
\ee
has the property that $\bra{\theta_-}\tau_3\ket{\theta_-}<0$, which obviously
violates the positivity condition of the squashing map.

Therefore we conclude that for the choice of the classical postprocessing we made there is
no complete positive squashing map onto target measurement given by Eq.~(\ref{sixstatetarget}).

\subsection{Alternative classical postprocessing and completely positive squashing map}\label{ssection:6statealternativeCPP}
As we mentioned in Section \ref{noisyCPPs}, the lack of complete positivity of the squashing map
is not always a big obstacle and can be overcome by introducing some additional noise to the
measurement data. This is done by choosing another CPP scheme. We will apply this trick for the
six-state active measurement device and show that the amount of noise one needs to introduce
is tolerable in QKD implementations.

We start by describing a new classical postprocessing.
This will be a mixture of the old postprocessing and the
completely noisy postprocessing that corresponds to the random assignment
of an outcome regardless of which basic event occurred (cf.~Eq.~(\ref{noisyscheme})).
For this type of postprocessing there is a positive squashing map $\Lambda_{\varrho_{fix}}$
with $\vr_{fix}=\eins_Q/2$ (cf.~Theorem \ref{mixedsquashmap} and Remark \ref{noisyfullrank}).

We use Eq.~(\ref{mixedChoi}) with $d_Q=2$ and $\vr_{fix}=\eins_Q/2$ to achieve the Choi matrix
for new squashing map
\be
\tau_{P,N,\textnormal{new}}(p)=(1-p)\tau_{P,N} + \frac{p}{4}\eins_Q\otimes\eins_{P,N},
\label{eq:sixstatePOVMnewold}
\ee
where $\tau_{P,N}$ is from Eq.~(\ref{ChoiMatrix6state}).

In order to check the positivity of $\tau_{N,\textnormal{new}}(p)$ we can again represent it
in the basis of Eq.~(\ref{6statebasis}) as a $12\times 12$ matrix of the parameters $N$ and $p$,
and show the positivity of its eigenvalues for $p\geq 1/3$. The proof is given in Appendix D.

Next we point out a connection between the parameter $p$ and an additional
penalty bit error rate $e$ in the six-state protocol, which
one needs to introduce in order to ensure the positivity of the squashing map.

\begin{rem}(Connection between white noise parameter and bit error rate)
The white noise parameter $p$ corresponds to the double bit error rate which
one needs to introduce as a penalty for a positive squashing map to exist. That is $p=2e$.
\end{rem}
{\it Proof:} The essential point for this is the fact that for the old full measurement
POVMs we have
\be
F^{(0,\alpha)}_{N,\textnormal{old}} + F^{(1,\alpha)}_{N,\textnormal{old}}=\eins_{N}.
\ee
Substituting this equation in Eq.~(\ref{eq:sixstatePOVMnewold}) will result in
\bea
F^{(0,\alpha)}_{N,\textnormal{new}}&=(1-p/2) F^{(0,\alpha)}_{N,\textnormal{old}} + \frac{p}{2} F^{(1,\alpha)}_{N,\textnormal{old}},\\
F^{(1,\alpha)}_{N,\textnormal{new}}&=(1-p/2) F^{(1,\alpha)}_{N,\textnormal{old}} + \frac{p}{2} F^{(0,\alpha)}_{N,\textnormal{old}}.\nonumber
\eea
This relates the full measurement POVM elements after additional noisy postprocessing to the elements before
the postprocessing. This relation concerns one particular measurement basis and can be interpreted as an additional bit flip
with probability $e=p/2$ after all double clicks have been assigned.
\qed

It follows from the last remark that we need to add $16.67\%$
of noise to our data in order for squashing map of the six-state
protocol to be completely positive.

To summarize we showed that the following statement holds.
\begin{thm}(Squashing model for six-state measurement with active basis choice)\label{thm:6stateactivesquasher}
There exists a squashing model with a qubit target measurement
for the six-state measurement with active basis choice (no matter how often
one chooses to measure in one of the three bases)
for the CPP scheme that randomly (with equal probability) assigns the double clicks
to single clicks and flips the single click bit values with probability 1/6.
\end{thm}

To conclude this section we point out that the recent results by Ma and L{\"u}tkenhaus \cite{ma12a}
allow us to say that this penalty error rate only needs to be
used to estimate the amount of privacy amplification necessary for the protocol.
As a matter of fact, we do not actually need to flip any bits and can therefore effectively reduce
the amount of information leaking to the eavesdropper.
In this case the squashing model can be
used for the protocol with one-way classical communication and one can provide a
secret key rate for an error rate up to $6.43\%$ if one uses the infinite-key-limit formula
$r=1-h(Q)-I_E(Q')$ (see Eq.~(A6) in Ref.~\cite{scarani09a})
with $Q'=(1-p_{\textnormal{flip}})Q+p_{\textnormal{flip}}(1-Q)$, where $Q$ is a bit error rate and
$p_{\textnormal{flip}}$ the flip probability from Theorem \ref{thm:6stateactivesquasher}.

\section{Extensions of squashing models. Biased active BB84 measurement.
Passive BB84 and six-state measurements.}\label{section:qubitextensions}

This section is devoted to several generalizations of the ideas that were laid out in
Sections \ref{linoptimpl} - \ref{ssection:6stateact}.
First we investigate a biased active BB84 measurement.
Then we turn our attention to the passive detection scheme (defined below) for the BB84 and six-state measurement devices.
Such devices are also often used in practical implementations of QKD protocols, which is often motivated
by the fact that a passive detection scheme requires fewer random bits and typically allows higher clock-rates (see e.g.~\cite{rarity92a,rarity94a}).
%

\subsection{Biased active BB84 measurement}\label{ssection:BB84biasedactive}

We start off by generalizing the squashing model for the BB84 measurement to
a device where the observer chooses the measurement basis according to classical
probabilities $p_+$ and $p_{\times}$ such that $p_{+}+p_{\times}=1$. That is, the
active ``at will'' choice of the observer is replaced by a random number generator.

This will result only in a coefficient in front of the POVM elements in Eq.~(\ref{sceffects0})
\bea
F_{\textnormal{vac}}=&\sum_{\alpha=+,\times}p_{\alpha}\ketbraa{0,0},\nonumber\\
F^{i,\alpha}_{\textnormal{sc}}=&p_{\alpha}\sum_{N=1}^{\infty}\ketbraaa{N}{i,\alpha}{},\\
F^{\alpha}_{\textnormal{dc}}=
&p_{\alpha}\sum_{N_0,N_1=1}^{\infty}\ketbraaa{N_{0},N_1}{\alpha}{},\nonumber
\eea
and will not affect the rest of the argument we made in Section \ref{ssection:BB84active}. Therefore we have the following theorem.
\begin{thm}(Squashing model for biased active BB84 measurement)
There exists a squashing model for the biased active BB84 measurement for which the basis
choice is made according to classical probabilities $p_+$ and $p_{\times}$ such that $p_{+}+p_{\times}=1$.
\end{thm}

\subsection{Passive detection scheme for the BB84 measurement}
In this section we will present the details for the passive BB84 measurement. The passive six-state
measurement can then be generalised straightforwardly.

\begin{figure}
\includegraphics[width=.99\columnwidth]{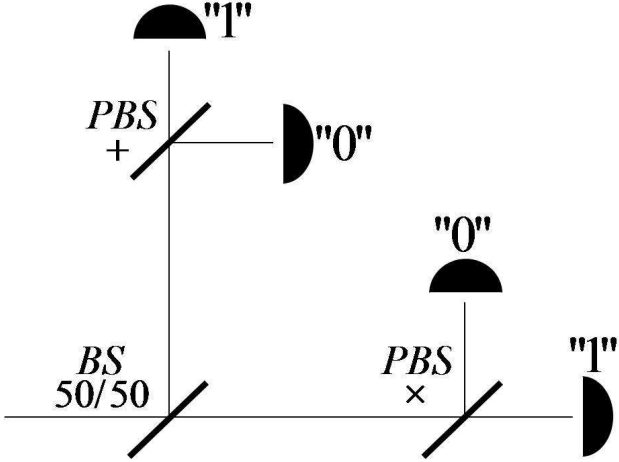}
\caption{The device for the passive BB84 measurement. Here it consists of two detection modules,
located at the output ports of a 50/50 beam splitter. Each detection module corresponds
to one of the polarization bases. \label{pasiveBB84}}
\end{figure}

In the passive BB84 measurement the observer uses a measurement device presented in Fig.~\ref{pasiveBB84}.
The whole measurement device consists of two detection modules that correspond to two detection bases.
Both detection modules are positioned at the two output ports of a 50/50 beam splitter.
This measurement outputs a bit value and a basis choice.

Interestingly, this type of detection scheme is more sensitive to the signals containing more than one photon.
Indeed, because of the $50/50$ beam splitter (cf.~Eq.~(\ref{5050bs}) in the appendix) the POVM elements will
take the form (see also \cite{nl99a})
\bea
F_{\textnormal{vac}}&=\halbe\sum_{\alpha=+,\times}\ketbraa{0,0},\nonumber\\
F^{i,\alpha}_{\textnormal{sc}}&=\sum_{N=1}^{\infty}\left(\halbe\right)^{N}\ketbraaa{N_{i}}{\alpha}{},\; i=0,1,
\label{sceffects}\\
F_{\textnormal{dc}}^{\alpha} &= \halbe\sum_{N_0,N_1=1}^{\infty}\ketbraaa{N_{0},N_1}{\alpha}{},\nonumber\\
F_{\textnormal{cc}} &= \halbe\eins +
\mathop{\sum_{\alpha=+,\times}}_{i=0,1}\sum_{N=1}^{\infty}\left(1-\frac{1}{2^N}\right)\ketbraaa{N}{i,\alpha}{},
\label{sceffects2}
\eea
where ``sc", ``dc" and ``cc" denote single clicks, double clicks within the same detection module, and
cross-clicks between different modules respectively.

\subsection{Relation between active and passive measurement devices via switching}\label{ssection:switching}
A closer comparison of Fig.~\ref{PBSBB84} and Fig.~\ref{pasiveBB84} reveals an important relationship
between the two detection schemes. First note that the single-click POVM elements are the same (up to an
$N$-dependent coefficient) for both detection schemes. As we will see shortly, this can
be crucial for the positivity of the squashing map. In fact, an active detection scheme can be generally represented as
a part of the passive detection scheme with a beam splitter which acts as a probabilistic classical switch.

\begin{figure}
\includegraphics[width=.99\columnwidth]{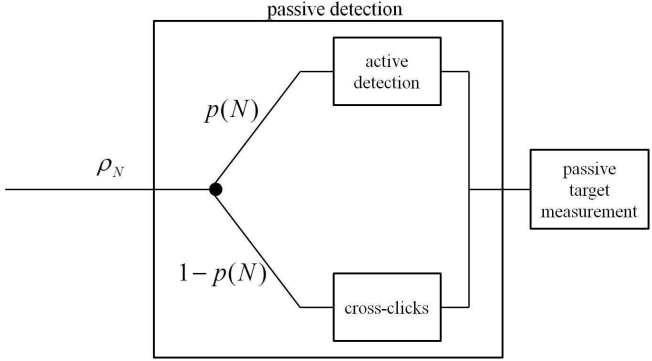}
\caption{Active detection scheme as a part of the passive detection scheme.
Any passive detection that is due to an input beam splitter can be thought of as
a switching between active detection and a cross-click. The probability
of switching depends only on the number of photons $N$ entering the
measurement device. \label{fig:switchingscheme}}
\end{figure}

Indeed, the initial beam splitter is only rerouting all photons in one direction
or the other direction or splitting-up the photons.
So we can think of it like a classical switch which either decides between the two polarization
bases like in an active scheme, or splits up the photons thus creating cross-clicks.
The probability for the first two cases are the same, so we can combine them into the problem
of active detection in Fig.~\ref{fig:switchingscheme}.
Therefore we can
think of the passive detection scheme as of a scheme consisting of two parts:
with probability $p(N)$ the system chooses to apply an active detection set-up,
and with probability $1-p(N)$ the system chooses to create a cross-click.
To give an example, for the passive BB84 measurement in
Fig.~\ref{pasiveBB84} we have $p(N)=\frac{1}{2^N}+\frac{1}{2^N}=\frac{1}{2^{N-1}}$.

As we will see now, this structure turns out to be very useful for the construction of squashing models and
especially for choosing the right CPP scheme for passive detection devices.

\subsection{Squashing model for the passive BB84 measurement}\label{ssection:BB846statepassive}
According to the previous section, we can rely on our knowledge of the
active detection scheme and consider the cross-click events separately.

As it turns out, since there exists a squashing
model for the active detection scheme (see Section \ref{ssection:BB84active}),
we can either discard all cross-clicks or assign them
to some bit value with some probability.

To be more precise, let us define two different CPP
schemes for the part of the squashing model that deals only with cross-clicks
(cf.~Fig.~\ref{fig:switchingscheme}) and hence leads to two different overall
squashing maps
\bea
\Lambda^{\textnormal{discard}}_{N,\textnormal{passive}} &= \frac{1}{2^{N-1}}\Lambda_{N,\textnormal{active}} +
\left(1-\frac{1}{2^{N-1}}\right)\Lambda_{\textnormal{vac}}\nonumber\\
\Lambda^{\textnormal{keep}}_{N,\textnormal{passive}} &= \frac{1}{2^{N-1}}\Lambda_{N,\textnormal{active}} +
\left(1-\frac{1}{2^{N-1}}\right)\Lambda_{\eins_Q/2},\label{eq:BB84switcher}
\eea
where the vacuum map $\Lambda_{\textnormal{vac}}$ disregards the input and forwards a
vacuum state to the target measurement (not to be confused with the vacuum flag (see the closing remark
in Section \ref{section:QND})).
We refer to $\Lambda^{\textnormal{discard}}_{N,\textnormal{passive}}$
as the squashing map corresponding to the overall postprocessing that discards all
cross-clicks and we refer to $\Lambda^{\textnormal{keep}}_{N,\textnormal{passive}}$ as the
squashing map that corresponds to the CPP scheme with random (in this case with
equal probabilities, because of the passive target measurement) assignment of cross-clicks.
Note that in principle we can choose the assignment of cross-clicks according to some
other (non-uniform) probability distribution. It will only mean that we would need
to change $\vr_{fix}$ in $\Lambda_{\vr_{fix}}$ accordingly,
but it will not affect the positivity of the squashing map, as long as $\vr_{fix}$
corresponds to a physical state.

Finally, the positivity of both passive maps in Eq.~(\ref{eq:BB84switcher}) for the BB84 case
follows from the positivity of the maps on the right hand side of Eq.~(\ref{eq:BB84switcher}),
which is in contrast to the six state measurement considered in the next section.

To summarize we have the following theorem.
\begin{thm}(Squashing model for the passive BB84 measurement)\label{thm:squasherBB84passive}
For the passive BB84 measurement there is a squashing model with a qubit target measurement no matter what the classical postprocessing
of the cross-clicks is, as long as there is a squashing model with a qubit target measurement for cross-click events.
\end{thm}

\subsection{Passive six-state measurement}\label{ssection:6statepass}
For the six-state measurement device with the passive detection scheme we can use the
same argument as we did in the last section for the BB84 measurement device. Here
however we cannot rely on the positivity of the squashing map for the active
part of the model because as we learned in Section \ref{ssection:6stateact} the
map is not completely positive. In virtue of this we cannot simply discard
all cross-clicks and we have to fall back to their random assignment.
We still will be using the map $\Lambda_{N,\textnormal{active}}$, although it is
not physical, as a tool to show the complete positivity of the overall map (see below).

Assuming that we assign cross-clicks with equal probabilities we have
the overall squashing map of the type
\be
\Lambda^{\textnormal{keep}}_{N,\textnormal{passive}} = \frac{1}{3^{N-1}}\Lambda_{N,\textnormal{active}} +
\left(1-\frac{1}{3^{N-1}}\right)\Lambda_{\eins_Q/2}.
\ee
The positivity of this squashing map will be, as per usually, investigated in terms of its
Choi matrix:
\be
\tau^{\textnormal{keep}}_{N,\textnormal{passive}}=\frac{1}{3^{N-1}}\tau_{N,\textnormal{active}} + \left(1-\frac{1}{3^{N-1}}\right)\frac{\eins_Q\otimes\eins_N}{4}.
\ee

Note that the chosen CPP scheme allows us to apply Reduction 2 here (Observation \ref{obs:PPorth})
and consider the Choi matrix only on the single-click subspace $P$: $\tau_{P,N,\textnormal{passive}}$.
The positivity of $\tau_{P,N,\textnormal{passive}}$
was discussed in Eq.~(\ref{eq:sixstatePOVMnewold}) where we concluded that
this matrix is positive whenever $p(N) = 1- 1/3^{N-1}$ and takes values $p(N) \in [1/3,1]$, which is
the case for any $N\geq 2$.

Therefore, all eigenvalues are positive
and we have shown the following.
\begin{thm}(Squashing model for the passive six-state measurement)
For the passive six-state measurement device there exists a squashing model if the classical
postprocessing randomly assigns (with equal probability) the double-clicks to a bit value
within the same basis where the double-click has occurred, and assigns the cross-clicks randomly (with
equal probability) to one of the possible bit values.
\end{thm}

\section{Passive multi-state qudit measurement device for prime-dimensional Hilbert spaces}\label{rubenswork}
In the previous section we discussed squashing models for measurement devices for which it was
self-evident to choose a qubit measurement as the target measurement. A possible way to generalize
the results of Sections \ref{ssection:BB84active} and \ref{ssection:6stateact}
is to consider a qudit measurement device as the target instead.
Here we present a general result for the passive detection scheme in the case of a qudit target measurement,
with $d$ being a prime number. Note that we only consider passive devices since this
generalization includes the six-state measurement device, for which an active choice does not work
(see Section \ref{ssection:6stateact}).

We will be setting the stage by recapitulating some known facts about mutually unbiased bases (MUBs)
and the reader who is familiar with this notion can skip the next section without losing the thread
of the paper.

\subsection{Background on mutually unbiased bases}
First we recall the definition of MUBs.
\begin{defn}(MUB)
Let $\{\ket{\psi_1}\dots\ket{\psi_d}\}$ and $\{\ket{\phi_1}\dots\ket{\phi_d}\}$ be two orthonormal bases in the
Hilbert space $\CC^d$. These bases are called {\bf mutually unbiased} if
\be
|\braket{\psi_i}{\phi_j}|=\frac{1}{\sqrt{d}}, \forall i,j
\ee
\end{defn}
The existence of a $d+1$ MUB, if $d$ is a prime number, was proven in Ref.~\cite{bandy08a}. The proof is constructive. Each basis
consists of the eigenvectors of
\be
Z_d, X_d, X_d Z_d, X_d \left(Z_d\right)^2,\dots, X_d \left(Z_d\right)^{d-1},
\label{MUBgenerators}
\ee
where $Z_d$ and $X_d$ are generalized Pauli matrices with the properties
\be
Z_d\ket{j}=w^j\ket{j},\: X_d\ket{j}=\ket{(j+1)\mbox{ mod } d},
\ee
where $\omega$ is the $d$-th root of unity.
The basis of $Z_d$ is referred to as the standard basis. It is possible to represent
all other matrices in Eq.~(\ref{MUBgenerators}) in terms of the standard basis. This
gives an explicit relationship between different bases representations
\bea
Z_d^{k}&= \sum_{i=0}^{d-1}\omega^{ik}\ketbraaa{1}{i,0}{},\nonumber\\
\left(X_d Z_d^{\alpha}\right)^{k}&= \sum_{i=0}^{d-1}\omega^{\alpha k
\left(i+\halbe(k - 1)\right)}\ketbraaa{1}{i+k,0,i}{}                
\label{eq:MUBrelationship}\\
&=\sum_{i=0}^{d-1}\omega^{ik}\ketbraaa{1}{i,\alpha+1}{},\nonumber
\eea
where $\alpha=0,\dots,d-1$.

One more fact that we will be using quite often is that the operators
\be
Z_d^{\alpha},(X_d Z_d^{\alpha})^{k}, \alpha=0,\dots,d-1,k=1,\dots,d-1
\label{quditbasis}
\ee
form a basis in the space of all operators acting on the Hilbert space of a qudit $\BB(\HH_d)$.

\subsection{Description of the measurement device: basic measurement}\label{measappMUB}

\begin{figure}
\includegraphics[width=.99\columnwidth]{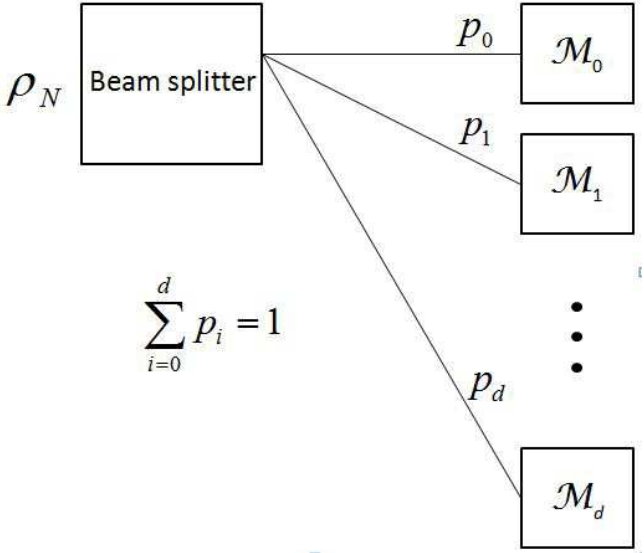}
\caption{Scheme of the measurement device for the qu$d$it QKD protocol
(where $d$ is a prime number). After passing the input beam splitter, which distributes
the signals into $d+1$ arms with probabilities $p_{\alpha}$ with $\alpha=0,\dots,d$, the input state is measured
in one of the $d+1$ mutually unbiased bases.
Each basis measurement contains $d$ different threshold photodetectors corresponding to
$d$ different polarization states within the basis $\alpha$. There are $d(d+1)$ polarization states in total.
\label{MUBapparatus}}
\end{figure}

The measurement device for the multi-state protocol is schematically represented in Fig \ref{MUBapparatus}.
It measures in $d+1$ different polarization bases. For each basis there are $d$ different polarization
states that are detected by one of $d$ threshold photodetectors in a corresponding detection module $\MM_{\alpha}$,
$\alpha=0,\dots,d$. Note that the described measurement device contains the passive six-state
measurement, if one assigns $p_{\alpha}=1/3$ for $\alpha\in\{x,y,z\}$.

From now on we will consider the situation where $p_{\alpha}=1/(d+1)$ for all $\alpha=0,\dots,d$ (possible generalizations
will be considered in the closing part, \ref{ssection:MUBconclusion}, of this section).

Since our measurement device contains threshold photodetectors we can perform a QND measurement
of the number of photons and split off the vacuum component first (see Observation \ref{obs:QNDreduction}). Hence we can
restrict ourselves to inputs containing $N\geq1$ photons.
In the $N$-photon subspace the basic POVM elements have the following form (see Appendix E for more details)
\bea
F_N^{i,\alpha}&=(d+1)^{-N}\ketbraaa{N}{i,0}{}\nonumber\\
F_{\textnormal{mc},N}^{\alpha}&=(d+1)^{-N}\left(\eins_{N}-\sum_{i=1}^d\ketbraaa{N}{i,0}{}\right)
\label{MUBBasicPOVM}\\
F_{\textnormal{cc},N} &= \left(1-\frac{1}{(d+1)^{N-1}}\right)\eins_N \nonumber,
\eea
with $i=1,\dots,d$ and $\ket{N}_{i,\alpha}$ is adopted to describe an event of the
detection of $N$ photons in the detector $i$ of the detection module $\MM_{\alpha}$.
From the structure of the basic POVM elements it is evident that we have to distinguish between multi-clicks (mc) $F_{\textnormal{multi},N}$ and cross-clicks (cc) $F_{\textnormal{cc},N}$.
The multi-clicks happen when different detectors within the same detection module (i.e.~the same basis choice) have a click
and is a generalization of a double-click for $d=2$.
Cross-clicks occur when several detectors in at least
two different measurements modules (two different bases) have a click.

\subsection{Target measurement, classical postprocessing and full measurement}\label{ssection:MUBtarget}

The target POVM elements can be easily deduced when we restrict to a single-photon input state (note the analogy to Eq.~(\ref{sixstatetarget}) for the
six-state measurement):
\bea
F_1^{i,\alpha}=\frac{1}{d+1}\ketbraaa{1}{i,0}{},\label{MUBtarget}\\ \alpha=0,\dots,d;\;i=0,\dots,d-1.\nonumber
\eea

It is not hard to see that the target POVM elements are single photon restrictions of the basic POVM elements in Eq.~(\ref{MUBBasicPOVM}). Therefore we choose a CPP scheme that does not affect single-click basic POVM elements.
Following the same lines as the BB84 and the six state squashing models, we choose the CPP scheme as follows:
\bi
\item Single clicks are mapped to the same single clicks
\item Multi-clicks are assigned equally randomly to one of the $d$it values in the same module $\MM_{\alpha}$.
\item Cross-clicks are assigned with probability $1/d(d+1)$ to one of the outcomes of the measurement
$\MM_{\alpha}$.
\ei

The chosen CPP scheme and Eq.~(\ref{MUBBasicPOVM}) imply the following form of the full POVM measurement elements
\be
\ftil_N^{i,\alpha}=F_N^{i,\alpha}+\frac{1}{d}F_{\textnormal{mc},N}^{\alpha} + \frac{1}{d(d+1)}F_{\textnormal{cc},N}.
\label{PPRuben}
\ee

\subsection{Positivity of the squashing map $\Lambda_N$}

To start off, we note that the chosen postprocessing scheme
allows us to use the results of
Section \ref{section:OrthoSubspace} and precede the squashing
map by a projection on the $d(d+1)$-dimensional space $P=\textnormal{span}\{\ket{N_{i}}_{\alpha}\}$, because
the projection on this space commutes with full measurement POVM elements (Observation \ref{obs:PPorth}).

All states in the orthogonal compliment $P_{\bot}$ will produce either multi-clicks or cross-clicks.
In this case the squashing map will output
a completely mixed qudit state $\eins_d/d$: $\Lambda_{P_{\bot},N}[\vr_N]=\eins_d/d$ for all $\vr_N$.
Therefore we have constructed a squashing map on the subspace $P_{\bot}$.


What is left is to construct is a completely positive $\Lambda_{P,N}$.
The adjoint of the squashing map must satisfy the linear constraints Eqns.~(\ref{probcond}), (\ref{probcondadjoint}):
\be
\Lambda^{\dagger}_{P,N}\left[\ftil_1^{i,\alpha}\right]=\ftil_{P,N}^{i,\alpha}.
\label{linconstrMUB}
\ee

We have the following result.
\begin{lem}(Complete positivity of $\Lambda_{P,N}^{\dagger}$)\label{MUBPositivity}
On the Hilbert space of interest there exists a completely positive map $\Lambda_{P,N}$ which fulfills the linear
constraints in Eq. (\ref{linconstrMUB}).
\end{lem}
{\it Proof:} The proof consists of two steps.
First, one has to construct a map $\Lambda^{\dagger}_{P,N}$ that fulfills the linear constraints. Second,
one needs to prove its complete positivity.
To begin with we note that Eq.~(\ref{linconstrMUB}) defines the map $\Lambda^{\dagger}_{P,N}$ on all target POVM elements
$\ftil_1^{i,\alpha}=F_1^{i,\alpha}$, which form a basis in $\BB(\HH_d)$.
This completeness is a starting point for the construction of the squashing map
$\Lambda_{P,N}$. As $\Lambda^{\dagger}_{P,N}$ is linear, this defines its action on any input operator.

In the second step we need to prove that the squashing map is completely positive. The proof is technical, and can be found in Appendix F.
\qed

This lemma finishes the construction of the squashing model for the qudit measurement device
with a uniformly distributing input beam-splitter.

\begin{thm}(Squashing model for the passive multi-state qudit measurement)\label{thm:fullMUBsquasher}
There exists a squashing model for the full ($d+1$ MUBs) passive multi-state qudit measurement device for
all prime numbers $d$.
\end{thm}

\subsection{Possible generalizations}\label{ssection:MUBconclusion}
In the concluding part of this section we make some remarks about possible ways to generalize the
results for the MUB measurement device.

First of all, one may want to allow for different input beam splitter ratios and choose
$p_{\alpha}\neq 1/(d+1)$. In the following remark we point out that it is not possible
if one wants to keep the CPP scheme of Section \ref{ssection:MUBtarget} unchanged,
i.e.~where the multi- and cross-clicks are uniformly distributed.

\begin{rem}\label{MUBCPPRemark}
For the CPP scheme chosen in Section \ref{ssection:MUBtarget} (uniform distribution of the multi- and cross-clicks)
the linear constraints Eq.~(\ref{linconstrMUB}) are fulfilled
if and only if the output probabilities of the input beam splitter in Fig.~\ref{MUBapparatus}
are all equal, i.e.~$p_{\alpha}=1/(d+1)$ for all $\alpha$.
\end{rem}
{\it Proof:} See Appendix G.

Note that this remark does not contradict the existence of a
squashing model for the passive BB84 measurement,
as the CPP scheme used there is not a special case of the CPP scheme of Section \ref{ssection:MUBtarget}.

The other possible extension of the result of Theorem \ref{thm:fullMUBsquasher} is to choose the input
beam splitter such that $p_{\alpha}$
are uniformly distributed over $k<d+1$ output arms and $p_{\alpha}=0$ for the rest of the beam
splitter outputs (this situation would include the BB84 measurement device). However we do not
have an analytic proof that a squashing model for such situation exists and leave it as an open question.

\section{The squashing model for time modes}\label{sectm}
So far we have provided examples in which squashing models take multiple photon signals to single photon
ones. However, there are other degrees of freedom in experimental measurements that
we have not accounted for in the squashing models considered so far.
For example, measurements
typically accept signals over a time window, whose responses from the measurement
(such as detector clicks) are grouped together into what is called an event.
During the time window of a detection event, it is possible that a measurement receives
signals in multiple time modes. In this section we address the question
of the existence of squashing models for measurement devices that accept multiple
time modes. First, we provide a squashing model for the multi-time-mode active
BB84 measurement device. Based on this result we prove the
existence of the squashing model for the multi-time-mode six-state
measurement. Finally, in accordance with Theorem \ref{thm:squasherBB84passive},
we apply the squashing model of the multi-time-mode active BB84 measurement
to the multi-time-mode passive BB84 measurement. Note that the following results will apply to any collection of spatial-temporal modes a detector might be
susceptible to, not only time modes.

\begin{figure}
\includegraphics[width=.99\columnwidth]{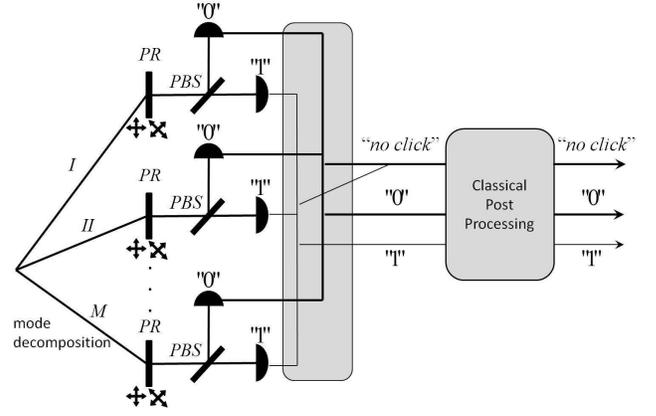}
\caption{Overall classical postprocessing for the multi-time mode squashing model. The overall vacuum or ``no-click'' event
is only registered if all single-time mode measurements output a ``no-click'' event.
If the click pattern in the single-time mode representation of the multi-time mode measurement contains only ``0''s (``1''s)
or vacuum, it is recorded as an overall ``0'' (``1''). Any other click pattern
is recorded as an overall double-click. The classical postprocessing randomly assigns the overall double-clicks to the value ``0'' or ``1'' in the corresponding basis with equal probability.
\label{modesquashCPP}}
\end{figure}

As we will show, a measurement device that receives an input in many time modes
can be thought of as many copies of that same measurement device,
each measuring with the same setting (for example, the
same basis) and each receives a single-time mode, followed by a suitable postprocessing
in order to combine the single-mode devices into the full device. Therefore, it is important to distinguish between two essential groups of basic events (see Fig.~\ref{modesquashCPP}):
\emph{single-time-mode events} and \emph{multi-time-mode events}. Note
that single detector clicks and multi-detector clicks can
correspond to a multi-time-mode event.

The situation seems to become rather cumbersome, because in general it is
unclear what classical postprocessing one should choose. However, if the
structure of the measurement device is such that it has a single-time-mode
squashing model for a specific classical postprocessing, then there
exists an overall classical postprocessing such that a multi-time-mode
squashing model exists. We have the following theorem.

\begin{thm}(Multi-time-mode squashing model for the active BB84 measurement)\label{thm:multitimeactiveBB84}
There exists a squashing model with a single-mode qubit target measurement for the multi-time-mode active BB84 measurement,
no matter how often one chooses to measure in either basis.
\end{thm}
\emph{Proof:} The proof relies on the fact that there exists a squashing model
for the single-time-mode device and in particular on the orthogonality of the
single-click and double-click subspaces (see Section \ref{ssection:activeBB84reduction}).

\emph{Classical postprocessing in the time domain.---}
The classical postprocessing
in the time domain is defined by grouping events that come from single-time-mode measurements.
If there is no click in any of the single-time-mode measurements we assign an ``overall'' vacuum to this event. We use ``overall'' to mean the output of this first step in the classical postprocessing.
If the outcome of every single-time-mode measurement is either ``0'' (``1'') or no-click we assign it to an overall ``0'' (``1''). To any other click pattern we assign an overall double-click.

These four types of events from a multi-mode measurement are then
forwarded to the classical postprocessing that allows a single-time-mode squashing
model.

In summary, the overall classical postprocessing for the multi-time-mode active BB84
measurement device is (cf.~Fig.~\ref{modesquashCPP}):
\bi
\item If there is no click in any of the single-mode devices then we call the
overall event a ``no-click''
\item If each single-mode device the event is  either ``0'' (``1'') or ``no-click'',
and there is at least one click event, we call the overall event ``0'' (``1'')
\item In the case of any other click pattern we call the event an ``overall double-click''
\item The overall single-clicks and vacuum events remain unchanged
\item The overall double-clicks are assigned with probability $1/2$
to ``0'' or ``1''
\ei

\emph{CP map for the classical postprocessing in the time domain.---}
In order to prove that there is a CP map
that preserves the structure of the incoming state and is compatible
with the provided classical postprocessing, we note that
according to Section \ref{section:QND} we can (without loss of generality)
perform a QND measurement of the photon number on each of the single-time modes.

\begin{figure}
\includegraphics[width=.99\columnwidth]{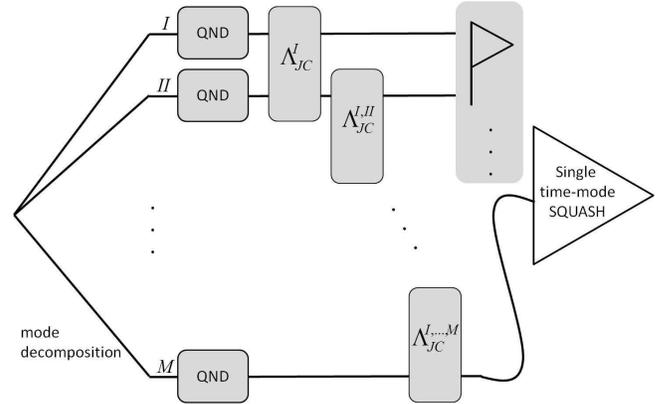}
\caption{The CP map compatible with the time-mode classical postprocessing. In the
first step, the number of photons is determined in every mode by performing a QND measurement.
Then a unitary map $U_{JC}^{k}$ is applied to the adjacent modes $k$ and $k+1$. This
map can be realized via a Jaynes-Cummings interaction, while the atomic degrees of freedom
are traced out afterwards. Last is a flag measurement performed on the first $M-1$
time modes. The state of mode $M$ is forwarded as input to the single-time-mode
squashing map. Depending on the outcome of the flag measurement either $\Lambda_P$
or $\Lambda_{P_{\bot}}$ is applied.
\label{modesquash}}
\end{figure}

Schematically the CP map on the overall system can be constructed by the composition of CP maps as
shown in Fig.~\ref{modesquash}. After the QND measurement
is done on each of the single-time modes, we combine all time modes into one single-time mode.
This is done via a unitary map $U_{JC}$ that depends on the outcome of the QND measurements. It
is applied to consecutive pairs of single-time modes (see Fig.~\ref{modesquash}).
The first single-time mode is then forwarded directly to a flag measurement (defined shortly),
whereas the second is an input to the next $U_{JC}$. This flag measurement
is a projection on the vacuum state of the first $M-1$ time modes and is performed
after the map $U_{JC}$ has been
applied on the $M-1$ and $M$ single-time modes. Depending on the outcome of this measurement, a particular
single-time-mode squashing map is applied to the state in time mode $M$.

To explain this in more detail, the map $U_{JC}$ can be described by using the Jaynes-Cummings
model. This map is unitary but depends on the total photon number in each mode (without
this knowledge it is impossible to perform the map). The crucial
property of this map is that it preserves the structure of the particular type of incoming
states independent of their polarization $\alpha=+,\times$ (cf. Appendix A in \cite{nl00a}):
\bea
&U^{I,II}_{JC}[\ketbraa{N_I,0}\otimes\ketbraa{N_{II},0}] \nonumber\\
&=\ketbraa{0,0}\otimes\ketbraa{N_I+N_{II},0}\nonumber,\\
&U^{I,II}_{JC}[\ketbraa{0,N_I}\otimes\ketbraa{0,N_{II}}] \nonumber\\
&=\ketbraa{0,0}\otimes\ketbraa{0,N_I+N_{II}}\label{timemodeCP}.
\eea

Thus we conclude that any state that produces an overall single-click will be
mapped to either one of four states:
\bea
&\ketbraa{0,0}^{\otimes(M-1)}\otimes\ketbraa{\sum_{k=1}^M N_k,0}\nonumber\\
&\ketbraa{0,0}^{\otimes(M-1)}\otimes\ketbraa{0,\sum_{k=1}^M N_k}\label{timemodesingleclicks}.
\eea

Therefore, if the outcome of the flag measurement is vacuum we know with certainty (due to
the unitarity of the map $U_{JC}$)
that the incoming multi-time mode state would have produced a single-click in
any single-time-mode measurement. If the outcome of the flag measurement is
not the vacuum then we know with certainty (again due to unitarity)
that the incoming state would
have produced an overall double-click in the multi-time-mode measurement.
By virtue of this, we proceed with the single-time-mode squashing map $\Lambda_P$ from
Section \ref{ssection:BB84active} if the flag signals vacuum, and $\Lambda_{P_{\bot}}$ otherwise.

Note that this procedure is compatible with the overall CPP scheme, since the CP map
in the time domain preserves the structure of the single-click subspace as it can be deduced
from Eqns.~(\ref{timemodeCP}) and (\ref{timemodesingleclicks}).

Since we are given that there is a squashing model for the single-time-mode measurement we
conclude the proof.
\qed

Note that the generalization of this result to other measurement devices is not straightforward.
On one hand, for qudit measurement devices, the first part of the proof holds
regardless of whether or not one measures in MUBs. Then
one needs the final single-mode squashing map to exist in order to claim the existence of the overall map.
On the other hand, for most of the passive devices, one would need to come up with a flag that can distinguish
between double- and cross-clicks, because CPP schemes for these devices usually distinguish
between these types of clicks (see Section \ref{ssection:BB846statepassive}).

Nonetheless there are two important corollaries from Theorem \ref{thm:multitimeactiveBB84}.
\begin{corol}
There is a squashing model with a qubit target measurement for the multi-mode active six-state measurement with noisy postprocessing.
\end{corol}
{\it Proof:} The proof repeats the proof of Theorem \ref{thm:multitimeactiveBB84}
until the point where one needs to apply
the single-time-mode squashing map. In this case the map from Section \ref{ssection:6statealternativeCPP}
is applied.
\qed

\begin{corol}
There is a squashing model with a qubit target measurement for the multi-mode passive BB84 measurement with a CPP scheme that
discards all cross-click events.
\end{corol}
{\it Proof:} Since all cross-click events are thrown away it is sufficient to have the
same flag as in the multi-mode active BB84 measurement. More precisely, the squashing map
is a sequence of three maps. First, all cross-clicks are projected onto a multi-mode vacuum state.
Second, the Jaynes-Cummings map is applied and a flag measurement is done. Third, depending on the
flag measurement, the squashing map for the single-time-mode active BB84 measurement is applied.
\qed


\section{Squashing model for the unbalanced phase-encoded BB84 (PE BB84) measurement device}
\label{Application}
\begin{figure}
\centering
\subfigure[]{
\includegraphics[width=.99\columnwidth]{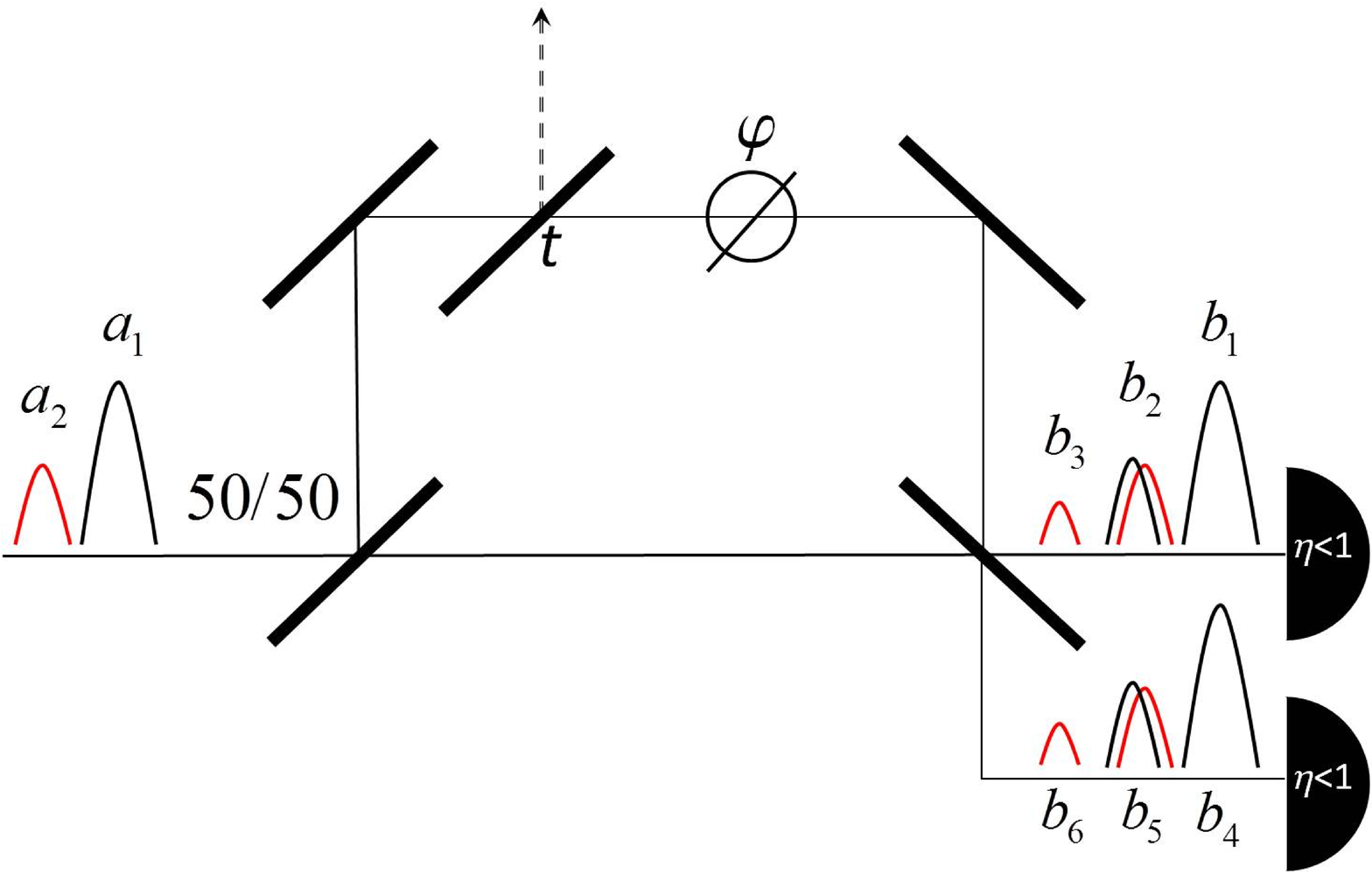}
\label{imperf}
}
\subfigure[]{
\includegraphics[width=.99\columnwidth]{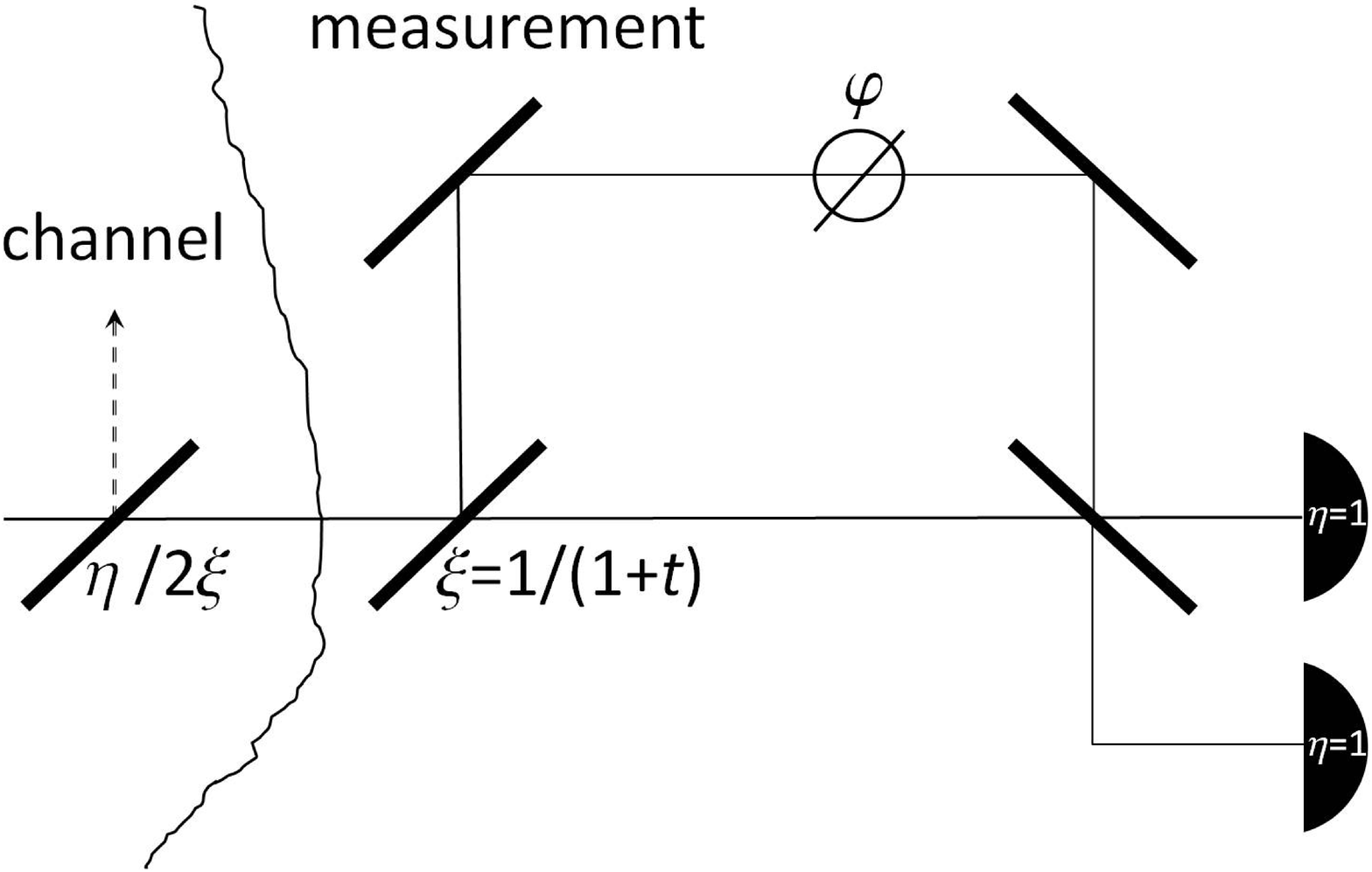}
\label{perf}
}
\caption{\label{PEBB84detector}(color online) The PE BB84 measurement device. (a) Model of the device
with imperfections: Two input modes $a$ carrying a relative phase $\phi_0=0,\pi/2,\pi,3\pi/2$
pass through a Mach-Zehnder interferometer with a lossy
phase-modulator in the longer arm (adjusted to either $\phi=0$ or $\phi=\pi/2$),
and finally give rise to 6 output modes $b$, detected by two threshold detectors of
equal efficiency $\eta$. (b) Equivalent model of the device with no imperfections but with some pre-detection loss,
which is absorbed in the channel and is modeled by a beam splitter with transmittance $\eta/2\xi$, and
an unbalanced input beam splitter $\xi=1/(1+t)$.}
\end{figure}

In the case of the unbalanced phase-encoded BB84 measurement device (PE BB84) the relevant information is always encoded
in the phase between two different time-modes that enter a Mach-Zehnder interferometer
(see Fig.~\ref{PEBB84detector}). This phase usually takes one of four values $\phi_0=0,\pi/2,\pi,3\pi/2$,
which is motivated by the corresponding QKD protocol (cf.~\cite{ferenczi12b}). The long arm of the interferometer has a lossy phase modulator, which can be adjusted to $\phi=0,\pi/2$ and whose loss is modelled
by a beam splitter with transmittance $t$. The signals pass through the interferometer and then
the modes are detected by two threshold detectors with equal efficiencies $\eta$ (see Fig.~\ref{imperf}).
Each of the two detectors can accept signals in three different time windows containing the modes
$(b_1,b_4)$, $(b_2,b_5)$ and $(b_3,b_6)$ respectively. Due to the structure of the device it is
clear that the relevant information about the phase is contained in the second time window,
since clicks in this time window correspond to the interference between the first
and the second input time modes.

The question of the existence of a squashing model for the unbalanced PE BB84 measurement is also practically motivated.
A change in the loss of the phase modulator changes the measurement operators.
For example, this implies that for the purpose of a security proof the lossless version of the device,
which is equivalent to the polarization BB84 measurement
(see Section \ref{ssection:PEBB84target} below), cannot be used anymore.

In order to not deal with losses and inefficient threshold detectors directly we can always consider
detection by a lossless interferometer with ideal threshold detectors but unbalanced input beam splitter
due to the loss in the phase modulator and additional pre-detection loss due to the originally
inefficient threshold detectors (cf.~\cite{ferenczi12b}). Such a measurement
device is presented in Fig.~\ref{perf}. Since we are not interested in any
pre-detection losses but solely in the mode of operation of the measurement device we ignore
the loss that is described by the beam splitter with transmittance $\eta/2\xi$ and consider
the ideal Mach-Zehnder interferometer with an unbalanced input beam splitter with transmittance $\xi$.

The POVM elements of the PE BB84 measurement device can be described by the mode operators $b_i$
that are related to mode operators $a_i$ of the incoming signal (cf.~\cite{ferenczi12b}).
The input-output relations (up to an unimportant overall phase) for the modes of interest are
\begin{align}
 \nonumber b_1&=\sqrt{\frac{\xi}{2}}a_1,\\
 \nonumber b_{2,\phi}&=\textnormal{e}^{-i\phi}\sqrt{\frac{(1-\xi)}{2}}a_1-\sqrt{\frac{\xi}{2}}a_2,\\
 \nonumber b_{3}&=\sqrt{\frac{(1-\xi)}{2}}a_2,\\
 \nonumber b_4&=\sqrt{\frac{\xi}{2}}a_1,\\
 \nonumber b_{5,\phi}&=\textnormal{e}^{-i\phi}\sqrt{\frac{(1-\xi)}{2}}a_1+\sqrt{\frac{\xi}{2}}a_2,\\
 b_{6}&=\sqrt{\frac{(1-\xi)}{2}}a_2.
 \label{PEBB84IO}
\end{align}

Having these relations in mind one can draw a connection to the usual polarization measurement. It is not
hard to see that the measurements in the middle time window for different settings of $\phi$
correspond to polarization measurements in two conjugate bases.
Moreover, if one combined clicks
in the first and in the third time window one would perform a measurement that corresponds to the measurement
in the standard basis in the polarization measurement device. Formally, for a lossless phase modulator ($t=1$)
we have the following POVM elements
\bea
F_Q^{2,0}&=\frac{1}{2}\ketbra{-},\;F_Q^{5,0}=\frac{1}{2}\ketbra{+}\label{PEBB84-passiveBB84-1}\\
F_Q^{2,\pi/2}&=\frac{1}{2}\ketbra{y_+},\;F_Q^{5,\pi/2}=\frac{1}{2}\ketbra{y_-} \label{PEBB84-passiveBB84}\\
F_Q^{1}+F_Q^{4}&=\frac{1}{2}\ketbra{0},\;F_Q^{3}+F_Q^{6}=\frac{1}{2}\ketbra{1}\label{PEBB84-passiveBB84-2},
\eea
where $F_Q^{T,\phi}$ is the POVM element on single-photon input in the time window $T$ and
for the phase modulator set to $\phi$.

Eqns.~(\ref{PEBB84-passiveBB84-1}), (\ref{PEBB84-passiveBB84}) establish the formal connection of the
lossless PE BB84 measurement (in the case where only events from the second time window are taken into account)
to the active polarization BB84 measurement.

\subsection{Target POVM elements and connection to the passive BB84 measurement device}\label{ssection:PEBB84target}
We choose the target measurement to be the full measurement restricted to the single-photon and vacuum input.
Since the useful events are detected in the second time window we group the target POVM
elements as follows

\begin{align}
F_Q^{2,\phi}&=\ketbraaa{1}{2,\phi}{},\nonumber\\
F_Q^{5,\phi}&=\ketbraaa{1}{5,\phi}{},\\
F_{\textnormal{out},Q}&=
\sum_{\stackrel{i=1,3,4,6}{\phi=0,\frac{\pi}{2}}}
\ketbraaa{1}{i,\phi}{}\nonumber.
\end{align}

The reason why we write the POVM elements here in terms of the output operators $b$ will be
more apparent in Section \ref{PEBB84BasicFullPOVMs}, where we consider $N$-photon input states.

In terms of the input states, which we denote by $\ket{0}=a_1^{\dagger}\ket{\textnormal{vac}}$
and $\ket{1}=a_2^{\dagger}\ket{\textnormal{vac}}$ respectively, the POVM element for the outside clicks can
be rewritten as
\be
F_{\textnormal{out},Q}=(1-\xi)\ketbra{0}+\xi\ketbra{1}=F_Q^{2,\phi}+F_Q^{5,\phi},\;
\forall \phi\label{PEBB84TargetRelations}.
\ee
This leads us to useful relations between the target POVM elements and the Pauli matrices
$\sigma_{\alpha}$, $\alpha=x,y,z$:
\bea
\frac{1}{\sqrt{\xi\left(1-\xi\right)}}\left(F_Q^{2,0}-F_Q^{5,0}\right)&=\sigma_x,\nonumber\\
\frac{1}{\sqrt{\xi\left(1-\xi\right)}}\left(F_Q^{2,\frac{\pi}{2}}-F_Q^{5,\frac{\pi}{2}}\right)&=\sigma_y,\nonumber\\
\frac{1}{2\xi-1}\left(2F_{\textnormal{out},Q}-\eins_2\right)&=\sigma_z.
\label{PEBB84usefulrels}
\eea

\subsection{Classical postprocessing scheme for the PE BB84 measurement device}\label{ssection:PEBB84CPP}
For a general input state with an undetermined number of photons in the input modes, the basic detection events can
be characterized by means of patterns of clicks on the two threshold detectors over the three time windows.
The total number of different events for each choice of the phase $\phi$ of the modulator is $2^6=64$.
As we demonstrated at the beginning of this section we can consider a lossless PE BB84
measurement device (cf.~Fig.~\ref{PEBB84detector}) by introducing some additional loss to the channel
and an unequal-ratio input beam splitter. It means that a "no-click" pattern never occurs if the incoming signal was not in a vacuum state. Therefore for two phase settings of interest $\phi=0$ and
$\phi=\pi/2$ we have 126 possible basic events.

In what follows we describe a basic event by a phase setting $\phi$ and a
click pattern $C:=\left(c_1,c_2,c_3,c_4,c_5,c_6\right)$, where each $c_i$ is either 0 or 1 (no click or click) and the index $i$ corresponds to the index of the output optical mode (see Fig.~\ref{PEBB84detector}).
This combination of indices provides an exact description of which detector has clicked and when.

Since our target measurement is a restriction of the basic measurement to the single photon subspace, we
are interested in a CPP scheme where single clicks are preserved and the rest of the postprocessing involves
only multiple clicks. This postprocessing needs to be valid in the terms that
were set in Section \ref{ssection:classPP}.

Using the linear dependency of the target POVM elements in Eq.~(\ref{PEBB84TargetRelations}) we performed
an exhaustive numerical search for a valid CPP scheme. There are many CPP schemes that are allowed by
the linear dependencies. Here we present the only postprocessing that,
as we will show shortly, allows a completely positive squashing map:

\begin{enumerate}
	\item Single clicks in either of the detectors in the second time window for either basis choice (i.e.~events with $C=\left(0,1,0,0,0,0\right)$ or $C=\left(0,0,0,0,1,0\right)$) are assigned to the corresponding single-photon events;
	\item Simultaneous clicks in the two detectors in only the second time window for either basis choice (i.e.~events with $C=\left(0,1,0,0,1,0\right)$) are assigned with probability $1/2$ to each of the single-photon measurement
outcomes for the same setting of the phase $\phi$;
	\item All events with clicks only in the first and the third time windows (outside clicks) are assigned to
an outside single-photon measurement event;
    \item Any event with clicks in both the second and any of the outer time windows is assigned
    with probability $1/2$ to the outside click event of the single-photon measurement event
    and with probability $1/8$ onto each of the four other events of the single-photon measurement event.
\end{enumerate}

\subsection{Basic and full measurement POVM elements}\label{PEBB84BasicFullPOVMs}
Since the PE BB84 measurement device contains threshold detectors we can use
the same argument as in Observation \ref{obs:QNDreduction}, Section \ref{section:QND} and consider the problem on
the $N$-photon subspace. From now on we will only consider $N$-photon input states.

According to the properties of the target measurement and the choice of the CPP scheme
we will distinguish between the following basic POVM elements:
\begin{align}
&F_N^{2,\phi}=\ketbraaa{N}{2,\phi}{},\nonumber\\
&F_N^{5,\phi}=\ketbraaa{N}{5,\phi}{},\nonumber\\
&F_{\textnormal{in,dc},N}^{\phi}=\sum_{k=1}^{N-1}\ketbraaa{k}{2,\phi}{}\otimes\ketbraaa{N-k}{5,\phi}{},\\
&F_{\textnormal{out},N}=\nonumber\\
&\sum_{\phi=0,\frac{\pi}{2}}\sum_{\stackrel{m_i\geq 0}{\sum_{i=1,3,4,6} m_i=N}}
\ketbraaa{m_1,m_3,m_4,m_6}{\phi}{},\nonumber\\
&F_{\textnormal{in,out},N}=\eins_N - F_{\textnormal{out},N} - \sum_{\phi=0,\frac{\pi}{2}}(F_{\textnormal{in,dc},N}^{\phi}+F_N^{2,\phi}+F_N^{5,\phi}).\nonumber
\end{align}
The last three POVM elements correspond to a
double click in the second time window for the particular
choice of the phase $\phi$ - $F_{\textnormal{in,dc},N}^{\phi}$,
to any click not in the second time window - $F_{\textnormal{out},N}$,
and to any cross-click between the second and one or both other time windows - $F_{\textnormal{in,out},N}$.

The complementarity condition gives us one more auxiliary POVM element
\be
F_{\textnormal{in},N}=\sum_{\phi=0,\frac{\pi}{2}}(F_{\textnormal{in,dc},N}^{\phi}+F_N^{2,\phi}+F_N^{5,\phi})
\ee
where $F_{\textnormal{in},N}$ is a POVM element corresponding to any click
in the second time window.

Note that neither $F_{\textnormal{in},N}$ nor $F_{\textnormal{out},N}$ depend on the
setting of the phase $\phi$ in the phase modulator. This can be seen if we write these elements in terms of
the incoming modes $a_1$ and $a_2$:
\bea
F_{\textnormal{in},N}&=\sum_{r=0}^N\xi^{N-r}\left(1-\xi\right)^r\ketbra{r,N-r},\label{PEBB84POVMsIO}\\
F_{\textnormal{out},N}&=\sum_{r=0}^N\xi^r\left(1-\xi\right)^{N-r}\ketbra{r,N-r},\nonumber
\eea
where $\ket{r,N-r}=\frac{1}{\sqrt{r!\left(N-r\right)!}}
\left(a^{\dagger}_1\right)^r\left(a^{\dagger}_2\right)^{N-r}\ket{\textnormal{vac}}$.

With this in mind we can write the full measurement POVM elements as
\bea
\ftil_N^{b,\phi}&=F_N^{b,\phi}+\frac{1}{2}F_{\textnormal{in,dc},N}^{\phi}+\frac{1}{8}F_{\textnormal{in,out},N},\nonumber\\
\ftil_{\textnormal{out},N}&=F_{\textnormal{out},N}+\frac{1}{2}F_{\textnormal{in,out},N},\label{PEBB84FullPOVMs}\\
\phi&=0,\frac{\pi}{2};\:b=2,5\nonumber.
\eea

\subsection{Positivity of the squashing map}\label{PEBB84SquashPos}
With full measurement elements provided in Eq.~\ref{PEBB84FullPOVMs} we have that the adjoint of
the squashing map $\Lambda_N$ has to fulfill the following linear constraints:
\bea
\Lambda^{\dagger}_N[F_Q^{b,\phi}]&=\ftil_N^{b,\phi},\phi=0,\frac{\pi}{2};\:b=2,5\nonumber\\
\Lambda^{\dagger}_N[F_{\textnormal{out},Q}]&=\ftil_{\textnormal{out},N}.
\label{PEBB84linconstr}
\eea

As in the previous examples, we investigate the positivity of the squashing map by investigating
the positivity of the corresponding Choi matrix $\tau_N=\eins\otimes\Lambda^{\dagger}_N[\ketbra{\psi^+}]$.
The Choi matrix will have no free parameters, since the target POVM elements can be seen
as an operator basis on the target Hilbert space. Using the decomposition of the
projector $\ketbra{\psi^+}$ in terms of Pauli matrices (see e.g.~Eq.~(\ref{eq:mesinpaulis}))
and employing the established connection between the Pauli matrices and the target
POVM elements (Eq.~(\ref{PEBB84usefulrels})) we have
\bea
&4\left(\eins\otimes\Lambda_N^{\dagger}\right)\left[\ketbra{\psi^+}\right]\nonumber\\
=&\eins_2\otimes\eins_N+\frac{1}{\sqrt{\xi\left(1-\xi\right)}}\sigma_x\otimes\left(\ftil_N^{2,0}-\ftil_N^{5,0}\right)\nonumber\\
&-\frac{1}{\sqrt{\xi\left(1-\xi\right)}}\sigma_y\otimes\left(\ftil_N^{2,\frac{\pi}{2}}-\ftil_N^{5,\frac{\pi}{2}}\right)\nonumber\\
&+\frac{1}{2\xi-1}\sigma_z\otimes\left(2\ftil_{\textnormal{out},N}-\eins_N\right)\nonumber\\
=&\eins_2\otimes\eins_N+\frac{1}{\sqrt{\xi\left(1-\xi\right)}}\sigma_x\otimes\left(F_N^{2,0}-F_N^{5,0}\right)\nonumber\\
&-\frac{1}{\sqrt{\xi\left(1-\xi\right)}}\sigma_y\otimes\left(F_N^{2,\frac{\pi}{2}}-F_N^{5,\frac{\pi}{2}}\right)\nonumber\\
&+\frac{1}{2\xi-1}\sigma_z\otimes\left(F_{\textnormal{out},N}-F_{\textnormal{in},N}\right).
\label{PEBB84Choi}
\eea
Here we used the classical postprocessing Eq.~(\ref{PEBB84FullPOVMs}) in order to write the
last equation in terms of the basic POVM elements.

When we write the basic POVM elements in terms of input mode operators $a_1,a_2$ (Eqns.~(\ref{PEBB84IO})),
we can show that $F_N^{b,\phi}=2^{-N}\ketbra{N_b}$ where $\ket{N_b}$ is a normalized vector.
Moreover from the same equations it follows that $|\braket{N_2}{N_5}|^2=(2\xi-1)^{2N}$.
Therefore the traceless operator $F_N^{2,\phi}-F_N^{5,\phi}=(\ketbra{N_2}-\ketbra{N_5})/2^N$ of rank 2
has eigenvalues $\pm\sqrt{1-(2\xi-1)^{2N}}$.

From Eq.~(\ref{PEBB84POVMsIO}) it follows that
\bea
&F_{\textnormal{out},N}-F_{\textnormal{in},N}=\label{eq:PEBB84BasicPOVMrelations}\\
&\sum_{r=0}^N\left(\frac{\xi^r}{(1-\xi)^{r-N}}-\frac{(1-\xi)^r}{\xi^{r-N}}\right)\ketbra{r,N-r}.\nonumber
\eea

With these relations in mind we can easily calculate the minimal eigenvalue of the Choi
matrix in Eq.~(\ref{PEBB84Choi}) numerically for any finite number $N$ of incoming photons.
We use Eq.~(\ref{eq:PEBB84BasicPOVMrelations}) in order
to represent the last term in Eq.~(\ref{PEBB84Choi}) as a $(N+1)\times(N+1)$ matrix.

\begin{figure}
\includegraphics[width=.99\columnwidth]{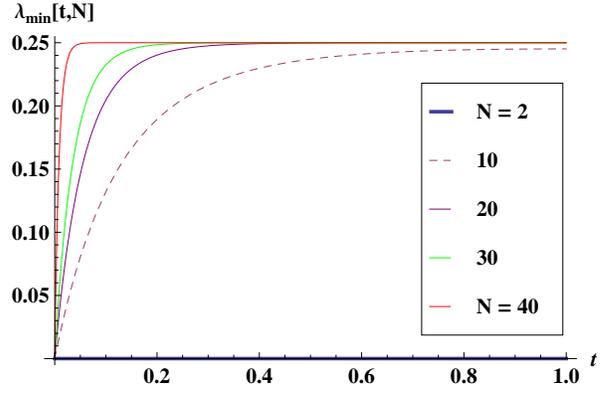}
\caption{\label{PEBB84mineig}(color online)
Minimum eigenvalue $\lambda_{\textnormal{min}}(t,N)$
of the Choi matrix $\tau_N$ of the adjoint squashing map $\Lambda_N^{\dagger}$ for the PE BB84
measurement device as a function of the loss $t$ in the phase modulator in the long arm
of the Mach-Zehnder interferometer. While for $N=2$ the minimal eigenvalue of $\tau_N$ stays
constantly zero, it becomes strictly positive for $N>2$ and any $t\in(0,1]$.}
\end{figure}

In Fig.~(\ref{PEBB84mineig})
we present the minimum eigenvalue of the Choi matrix as a function of the loss
in the phase modulator $t$ (recall that $\xi=1/(1+t)$) for different numbers of incoming photons $N$. One can readily see
that for any $t\in(0,1]$, $\lambda_{\textnormal{min}}(t,N)$ is non-negative. In fact, the minimum
eigenvalue of the Choi matrix is strictly positive for any $N>2$.
Moreover, it is clear from our numerics that $\lambda_{\textnormal{min}}(t,N)$ is
a nondecreasing sequence in $N$ for any $t\in(0,1]$, i.e.~$\lambda_{\textnormal{min}}(t,N+1)\geq\lambda_{\textnormal{min}}(t,N)$ for
all $t\in(0,1]$.

We will underpin our numerical findings by investigating asymptotical behavior of
$\lambda_{\textnormal{min}}(t,N)$ analytically. We derive a lower bound on the minimal
eigenvalue of the Choi matrix $\tau_N$ in two steps. First, as in the proof of Theorem \ref{mixedsquashmap},
we again use Weyl's inequalities in order to derive a lower bound on the
sum of the Hermitian matrices in Eq.~(\ref{PEBB84Choi}).
Second, we use the fact that $\lambda_{\textnormal{min}}(A\otimes B)\geq
\lambda_{\textnormal{min}}(A)\lambda_{\textnormal{min}}(B)$.
After some algebra the bound for all $N$ is given by
\bea
4\lambda_{\min}[t,N]\geq& 1-\frac{2(1+t)}{2^N\sqrt{t}}\sqrt{1-\left(\frac{1-t}{1+t}\right)^{2N}}\\
&-\left(\frac{\frac{1}{(1+t)^{N-1}}-\frac{t^N}{(1+t)^{N-1}}}{1-t}\right)=:f_N(t)\nonumber.
\eea

The function $f_N(t)$ is a monotonically increasing sequence of $N$ for any $t\in(0,1]$, i.e.~$f_{N+1}(t)-f_N(t)\geq 0$.
Moreover it holds that $\lim_{N\rightarrow\infty}f_N(t)=1$. Therefore for any $\epsilon$
there exists an $N_0$ such that for any $N>N_0$, $|f_N(t)-1|<\epsilon$.
Since $N_0$ is finite and for any finite $N_0$ one can calculate the minimum positive eigenvalue of the
Choi matrix explicitly, it follows that for all $N>N_0$
\be
4\lambda_{\textnormal{min}}(t,N)\geq 1-\epsilon\geq 0,
\ee
which concludes the proof of the positivity of the adjoint of the squashing map $\Lambda^{\dagger}$ and
implies the existence of the squashing model for the PE BB84 measurement device with a lossy
phase modulator.

\section{Application of squashing models to QKD protocols}\label{section:QKDapps}

In the section we shortly explain the application of squashing models for QKD. We start with the standard usage, followed by an advanced application, first put forward in Ref.~\cite{ma12a}, which provides slightly better key rates for various protocols. In the end we exemplify their difference for a BB84 protocol suffering from additional double click events on the receiver's side.

We  demonstrate the standard usage of squashing models in the security analysis of QKD using entanglement based QKD protocols, but all results straightforwardly apply also to prepare-and-measure schemes that do not use physical entanglement.

Let us start with a quick review of the essentials about QKD: An entanglement-based QKD protocol is realized in two phases: in a first phase, the quantum phase,  the two legitimate parties Alice and Bob perform measurements on their share of a tri-partite pure states, which can be thought of as being prepared by the adversary, Eve. In a subsequent second phase, the classical communication phase, Alice and Bob use an authenticated public channel to create a shared secret key from their data. The classical communication phase includes, typically, a sampling of their joinly correlated data that have been created in the quantum phase, a key map, error correction, and privacy amplification. The key map fixes which part of the data become key material. In our formulation we call these data $X$ and we assume without loss of generality that they are in Alice's hand. In error correction, Alice and Bob exchange additional information about $X$ using their available data to make sure that also Bob has a copy of the key material. At this stage, the adversary could still be correlated with $X$, thus having information about it. Privacy amplification turns the key material $X$ into a secret key $K$ by applying some privacy amplification function out of a pre-defined set of functions. The resulting key can be shown to be arbitrarily close to a perfectly secure key. For our purpose, we do not need to deal with the exact security statement, or with finite size effects in QKD. Instead we deal only with the secret key rate in what is known as the {\em infinite-key limit}.

In the security analysis we can calculate the guaranteed achievable secret key rate $R$ as number of secret key bits per bit of key material $X$ as
\begin{equation}
\label{eq:generic_rate}
R := \left(\inf_{\rho_{ABE} \in \Gamma_{ABE}} S(X|E)\right) - \delta_{\rm leak}, \;
\end{equation}
where  $S(X|E)$ is the conditional von Neumann entropy of Eve on the key material $X$, minimized over a set of potential underlying states $\rho_{ABE}$ (which can be thought of without loss of generality as pure states). This set is  constrained by Alice's and Bob's observations during sampling in the classical communication phase.  Each state $\rho_{ABE}$ represents an eavesdropping strategy which leads to particular conditional states $\rho_E^x$ in Eve's hand conditioned on an element $x$ of the key material $X$. These states summarize that part of Eve's knowledge about the key material  which stems from her preparation of (or interaction with) the physical systems on which Alice and Bob perform their measurements. The von Neumann entropy $S(X|E)$ is a function of these states and the probability of occurrence of the elements of $X$, as influenced by the public communication protocol. The last term, $\delta_{\rm leak}$ is the information content about the key material $X$ that leaks to Eve during the classical communication phase, including, for example, the number of bits per key material that have been announced publicly during error correction.

Squashing models can now be used to simplify the minimization calculation in the key rate expression. This is of particular importance for optical implementations of QKD where, for example,  Bob performs some type of photon-counting measurements, which need to be described on the infinite-dimensional Hilbert space of optical modes. Thus the set $\Gamma_{AB_FE}$ can have a rather complex form: it is given by all pure tri-partite quantum states $\rho_{ABE}$ which satisfy the condition that they give the observed correlations of data for Alice and Bob (including post-processing on Bob's side).

\begin{figure}
\centering
\subfigure[]{
\includegraphics[width=.99\columnwidth]{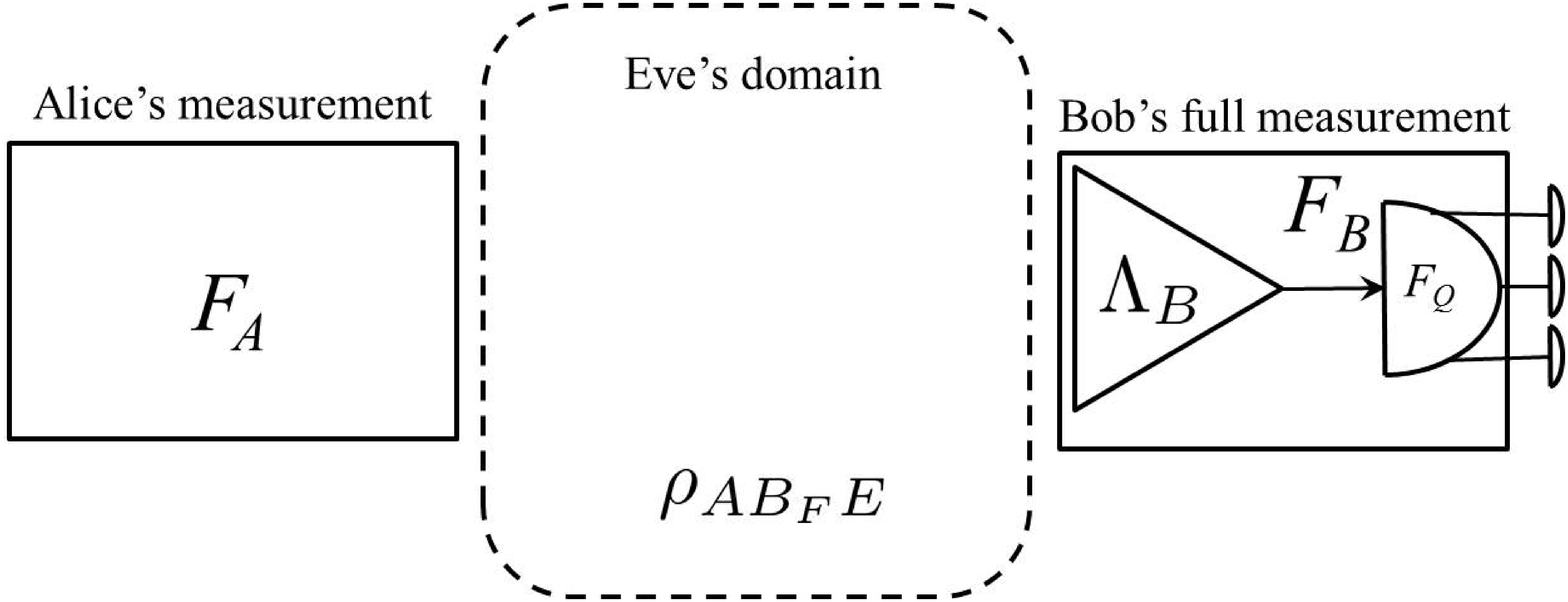}
}
\subfigure[]{
\includegraphics[width=.99\columnwidth]{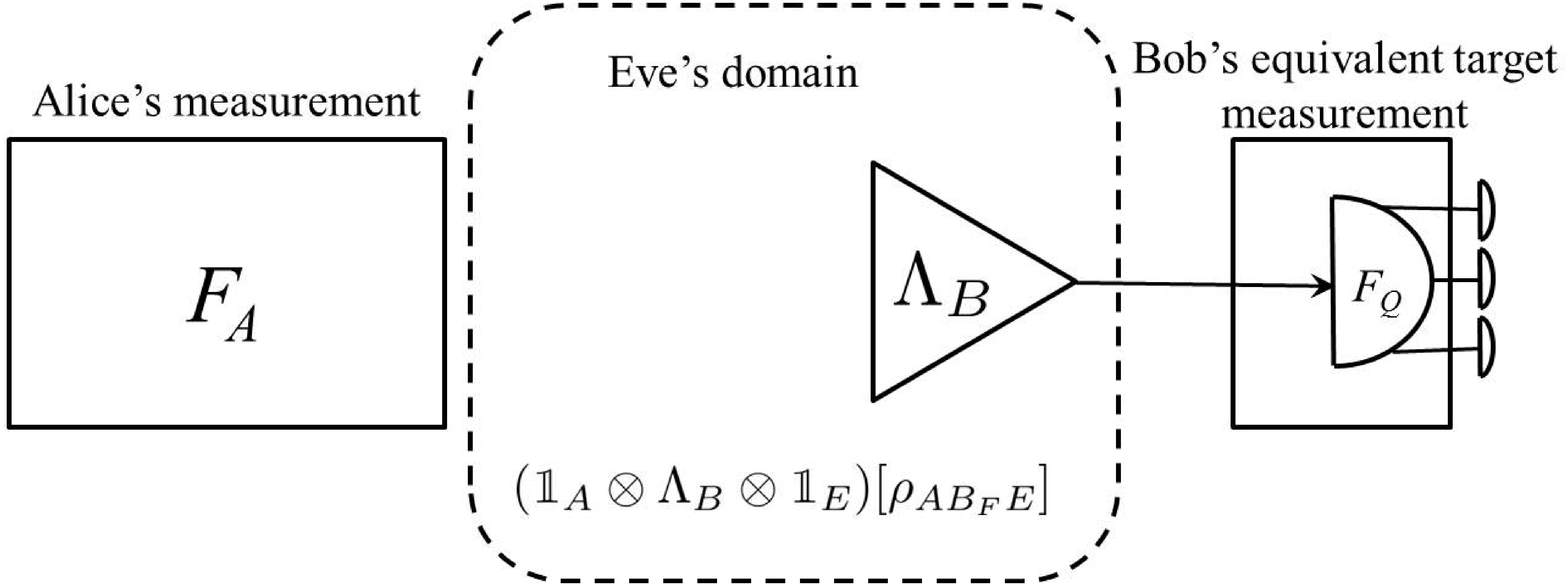}
}
\subfigure[]{
\includegraphics[width=.99\columnwidth]{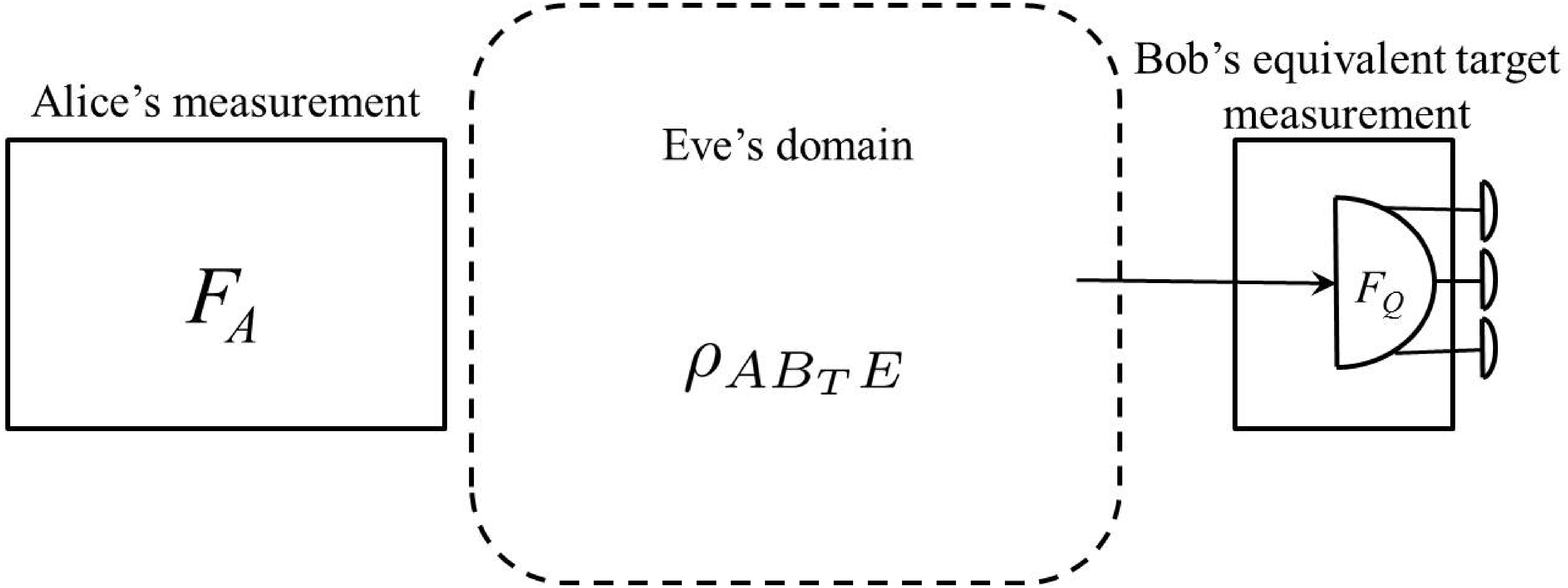}
}
\caption{\label{fig:squash_qkd} Explanation of the standard squashing models application in QKD: Case (a) represents a general attack against the full measurement scheme, while case (b) shows the reformulation of the same attack using the squashing model which results from enforcing the application of the squashing map on system $B$ of the general attack. Case (c) then represents a general attack of the scheme using the target measurement, which now clearly contains the case (b), and thus case (a). Thus the key rate of case (c) is a lower bound of the key rate in case (a), as stated in Theorem \ref{thm:generalQKD}.}
\end{figure}

\begin{thm} \label{thm:generalQKD}
The key rate $R_F$ using the full measurement can be lower bounded by a key rate using the target measurement, $R_T$ as
\begin{equation}
\label{eq:squash_rate}
R_F \geq R_T:=  \left(\inf_{\rho_{AB_TE} \in \Gamma_{AB_TE}} S(X|E)\right) - \delta_{\rm leak}, \;
\end{equation}
where $\rho_{AB_TE}$ has now the lower dimensional target system $B_T$ as one component. The set $\Gamma_{AB_TE}$ contains all such density matrices which are again constrained by the observed correlations for Alice and Bob. These constraints can now be formulated using the target measurements.
\end{thm}
The same statement can be made for the measurement on system $A$ if that measurement admits a squashing model.
To verify the theorem, we can follow a simple argument which is illustrated in Fig.~\ref{fig:squash_qkd}. As a first step note that an eavesdropping strategy against the full measurement protocol, as represented by a state $\rho_{AB_FE}$ and shown in Fig. \ref{fig:squash_qkd} (a), is now equivalent to an eavesdropping strategy described  by the state $\left(\openone_{A} \otimes \Lambda_B  \otimes \openone_E\right)[\rho_{AB_FE}]$, as the squashing map is applied to system $B_F$ without loss of generality before the target measurement is being performed, as shown in Fig.~\ref{fig:squash_qkd} (b). In the next step, we enlarge the set of possible eavesdropping strategies by directly admitting all density matrices of the type $\rho_{AB_TE}$ that are compatible with the observations, thereby dropping the constraint that it must be obtained by applying the squashing map $\Lambda$ to the full system (see Fig.~\ref{fig:squash_qkd} (c)). As a result of enlarging the set of allowed eavesdropping strategies over which the infimum in Eq.~(\ref{eq:generic_rate}) is taken, the resulting key rate can only decrease.

Note that the above theorem applies to the full measurements, that is, Bob performs the post-processing on his basic measurement results. Next, we draw the attention to the fact that the above key rate can be improved, as shown by Ma and L{\"u}tkenhaus in Ref. \cite{ma12a}. The motivation for this improvement comes from the fact that the postprocessing of the basic measurements helps to establish a squashing model to simplify the evaluation of the  conditional entropy $S(X|E)$ over all eavesdropping strategies. However, the same post-processing, typically,  increases the amount of communication  required during error correction from $\delta_{leak}^{\rm basic}$ to $\delta_{leak}\geq \delta_{leak}^{\rm basic}$ . The key observation in Ref. \cite{ma12a} is that these two points can be separated: we are allowed to use post-processed data to formulate constraints on $\rho_{AB_TE}$ to evaluate $S(X|E)$, but we do not actually need to apply the post-processing to our data, as long as the key material variables $X$ are not affected by the post-processing.
Formally, we can states this as

\begin{thm}\label{thm:QKDimproved}
For a QKD scheme, which uses basic measurements, post-processing of the basic measurements that leads to full measurements and allows for squashing model with some target measurement, we can find the key rate
\begin{align}\label{eq:helper2}
R_{AB_T}^{imp} \geq \left(\inf_{\rho_{AB_TE} \in \Gamma_{AB_TE}} S(X|E)\right) - \delta_{\rm leak}^{\rm basic}.
\end{align}
This key rate holds as long as the key material $X$ is unaffected by the post-processing. It uses the squashing model to estimate the term involving $S(X|E)$, but shows a reduced value of the term $\delta_{\rm leak}^{basic}$ which represents the amount of information on $X$ that leaks during the error correction phase of QKD based on Bob's basic measurement events.
\end{thm}

To illustrate the effect,  consider an example of the BB84  and compare the key rates when one applies either Theorem \ref{thm:generalQKD} or Theorem \ref{thm:QKDimproved}.
If we assume an initial symmetric data behavior of the type
\begin{equation}
\label{eq:observed_data}
P(X,Y) \!=\! \left\{
\begin{array}{l}
P(0,0)=P(1,1)=\frac{1}{2}P_{\rm single}(1-e) \\
P(0,1)=P(1,0)=\frac{1}{2}P_{\rm single}e \\
P(0,d)=P(1,d)=\frac{1}{2}(1-P_{\rm single}), \end{array} \right.
\end{equation}
with $P_{\rm single}$ being the probability to obtain a single-click event and $e$ as the error rate within the single-click events. Post-processing that allows a squashing model for these measurements involves a random assignment of double-clicks to the values $0$ and $1$, which  will raise the error rate from $e$ to $e_{PP}:=P_{\rm single} e+\frac{1}{2}(1-P_{\rm single})$.
Using the squashing model, we find
$$\inf_{\rho_{AB_TE} \in \Gamma_{AB_TE}} S(X|E) \geq 1-h[e_{PP}],$$
where $h[x]= - x \log_2 x - (1-x) \log_2(1-x)$ is the binary entropy function.

Now let us have a look at the amount of error correction we have to do, as quantified by $\delta_{\rm leak}$ (with post-processing) and $\delta_{\rm leak}^{\rm basic}$ (without post-processing).

Assuming that we can reach the Shannon limit for error correction, we find  $\delta_{\rm leak}= h(e_{PP})$, while we have  $\delta_{\rm leak}^{basic}=h[(1-P_{\rm single})+ P_{\rm single}h_2(e)]$.  Consequently, we find for the key rate according to Theorem  \ref{thm:generalQKD}
$$R_{AB_T} = 1-2 h[e_{PP}],$$
while we can find as improved key rate according to Theorem \ref{thm:QKDimproved}
$$R_{AB_T}^{imp} = 1-h[e_{PP}] - h[(1-P_{\rm single})+ P_{\rm single}h_2(e)],$$
which is a strict improvement of the rate due to the concavity of the binary entropy function.

\section{Conclusion and Outlook}
In this work we have further developed the ideas of Ref.~\cite{beaudry08a}. We gave a rigorous definition of
the squashing model and precisely defined the role of classical postprocessing in this setting.
In summary, the squashing models give us a tool for the truncation of Hilbert spaces under adversarial
conditions. More precisely, in the context of quantum communication in the presence of an adversary that can tamper
with the transmitted signals by employing their high dimensionality, it is equivalent to the scenario where one only performs the target measurement instead of a high dimensional measurement. This undoes the adversary's advantage, because there is no information that can be gained
by making a measurement on the high-dimensional system above what can be gained from the target measurement on the
lower dimensional system. Thus, the squashing models can be effectively applied, for example, in the optical implementations
of QKD or coin-tossing.

For QKD applications we constructed squashing models
for various types of measurement devices, which are used in optical implementations of corresponding QKD protocols.
For instance, this implies that a security proof of these protocols can be generalized from single photons to the multi-photon case.

Several generalizations can be made for results presented in this work. First of all, it would be interesting to generalize
the results on the qudit measurement to the case where the input beam splitter has less than $d+1$ output arms.
Our intuition
strongly suggests that the squashing model might exist if the number of output arms is strictly less than $d+1$.
This is due to the fact that removing the output arms would effectively introduce
free parameters in the Choi matrix of the squashing map, which is usually enough to guarantee the complete positivity
of the corresponding squashing map (cf.~active BB84 and six-state measurements). However these findings would heavily
rely on numerical findings. It would be therefore desirable to find some type of symmetry argument that can be used to tackle
this problem analytically.

\section*{Acknowledgements}
This research has been supported by NSERC (via the Discovery Program and the Strategic Project Grant FREQUENCY) and  the Ontario Research Fund (ORF). Oleg Gittsovich is grateful for the support of the Austrian Science Fund (FWF) and Marie Curie Actions (Erwin Schr\"odinger Stipendium J3312-N27). We would like to thank Marcos Curty, Eleni Diamant, Hoi-Kwong Lo, Momtchil Peev for many enlightening discussions.

\setcounter{thm}{0} \setcounter{equation}{0} \setcounter{figure}{0}
\renewcommand{\theequation}{A-\arabic{equation}}
\renewcommand{\thefigure}{A-\arabic{figure}}
\renewcommand{\thethm}{A-\arabic{thm}}

\section*{{\bf APPENDIX A}}\label{AppA}
In this appendix we summarize some known facts about the natural representation of
super-operators. Further details on this topic can be found in \cite{bengtsson06a}.
\begin{rem}(Equivalence of the natural and Choi-Jamio{\l}kowski representations)
Let $\Theta$ be a super-operator $\Theta(X)=Y$, with $X\in L\left(\HH_A\right)$ and $Y\in L\left(\HH_B\right)$ being linear operators. Let $\{\ket{e_m}\}$
and $\{\ket{f_{\mu}}\}$ denote orthonormal bases in $\HH_A$ and $\HH_B$ respectively. On the one hand
the natural representation of $\Theta$ is defined as a map
\be
\Gamma_\Theta : \vek{X}\mapsto\vek{\Theta(X)},
\ee
where $\vek{X}$ denotes column-wise vectorization of matrix $X$.
$\Gamma_\Theta$ is a linear operator $\Gamma_\Theta\in L\left(\HH_A\otimes\HH_A,\HH_B\otimes\HH_B\right)$
which can be represented by a matrix
\be
\Gamma_{\Theta_{\stackrel{\mu m}{\nu n}}}=\bra{f_{\mu},f_{\nu}}\Gamma_{\Theta}\ket{e_m,e_n}.
\ee
On the other hand the Choi-Jamio{\l}kowski representation is defined by the map \cite{jamiolkowski72a,choi75a}
\be
\tau_{\Theta} : L\left(L\left(\HH_A\right),L\left(\HH_B\right)\right) \rightarrow L\left(\HH_A\otimes \HH_B\right)
\ee
and can be represented by a matrix
\be
\tau_{\Theta_{\stackrel{n m}{\nu \mu}}}=\bra{e_n,f_{\nu}}\tau_{\Theta}\ket{e_m,f_{\mu}}.
\ee
One can readily see that $\Gamma_{\Theta}$ is a reshuffled version of the $\tau_{\Theta}$
\be
\bra{f_{\mu},f_{\nu}}\tau^R_{\Theta}\ket{e_m,e_n}=\bra{e_n,f_{\nu}}\tau_{\Theta}\ket{e_m,f_{\mu}}.
\ee
\end{rem}

\setcounter{thm}{0} \setcounter{equation}{0} \setcounter{figure}{0}
\renewcommand{\theequation}{B-\arabic{equation}}
\renewcommand{\thefigure}{B-\arabic{figure}}
\renewcommand{\thethm}{B-\arabic{thm}}

\section*{{\bf APPENDIX B}}
\begin{figure}
\includegraphics[width=0.99\columnwidth]{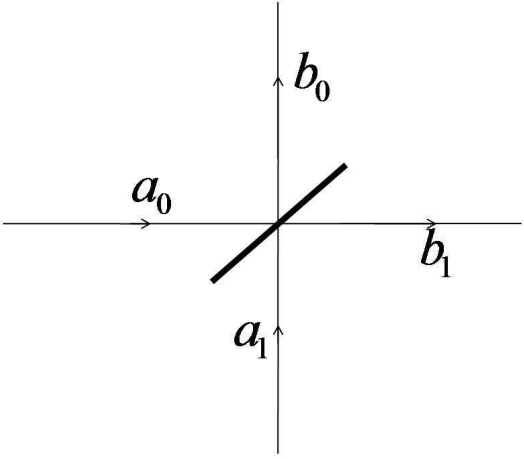}
\caption{Beam splitter presented as a four port device with $a_0$ and $a_1$, $b_0$
and $b_1$ being the input and output modes respectively. \label{BS1}}
\end{figure}

Beam splitters are well studied objects in quantum optics \cite{yurke_su2_1986,ou_relation_1987}. They are considered as
four port devices with two input and two output ports (Fig.~\ref{BS1}).

For a beam splitter with transmittance $T$ and reflectivity $R$ ($R+T=1$) the following relations
between the input and output modes hold
\bea
b_0&=\left(\sqrt{T}a_1 + \mbox{e}^{i\alpha}\sqrt{R}a_0\right)\nonumber\\
b_1&=\left(\sqrt{T}a_0 + \mbox{e}^{i(\pi-\alpha)}\sqrt{R}a_1\right),
\label{BSrels}
\eea
provided an $\alpha$ phase shift between the reflected and transmitted beams.
Without loss of generality, one can set $\alpha=0$.

Both the transmittance and the reflectivity of the beam splitter may depend on the frequency,
direction of propagation, and on the polarization of the incident light. For usual $50/50$ beam
splitter the relations simplify to
\bea
b_0&=\frac{1}{\sqrt{2}}\left(a_0 + a_1\right)\nonumber\\
b_1&=\frac{1}{\sqrt{2}}\left(a_0 - a_1\right)\label{5050bs}.
\eea


\setcounter{thm}{0} \setcounter{equation}{0} \setcounter{figure}{0}
\renewcommand{\theequation}{C-\arabic{equation}}
\renewcommand{\thefigure}{C-\arabic{figure}}
\renewcommand{\thethm}{C-\arabic{thm}}

\section*{{\bf APPENDIX C}}
\label{appendA}
Here we give explicit formulas for the matrix representations of the Choi matrices used for the squashing map in the BB84 active basis measurement's squashing model: $M(\tau_{P,N,\textnormal{fix}})$ and $M(\tau_{P,N,\textnormal{open}})$. We also discuss how to solve $M(\tau)\geq 0$ to ensure the complete positivity of the squashing map. The Choi matrices are given by
\begin{widetext}
\bea
M(F^{(0,z)}_{P,N}-F^{(1,z)}_{P,N})\equiv M_z &=
\left(\begin{array}{cccc}
1 & 0 & s & s\\
0 & -1 & -s & (-1)^{N+1}s\\
s & -s & 0 & \left(1-(-1)^{N}\right)s^2\\
s & (-1)^{N+1}s & \left(1-(-1)^{N}\right)s^2 & 0
\end{array}\right) \nonumber\\
M(F^{(0,x)}_{P,N}-F^{(1,x)}_{P,N})\equiv M_x &=
\left(\begin{array}{cccc}
0 & \left(1-(-1)^{N}\right)s^2 & s & -s\\
\left(1-(-1)^{N}\right)s^2 & 0 & s & (-1)^{N+1}s\\
s & s & 1 & 0\\
-s & (-1)^{N+1}s & 0 & -1
\end{array}\right) \label{BB84exact}\\
M(\eins_M)&=
\left(\begin{array}{cccc}
1 & 0 & s & s\\
0 & 1 & s & (-1)^{N}s\\
s & s & 1 & 0\\
s & (-1)^{N}s & 0 & 1
\end{array}\right) \nonumber\\
M\left((\Lambda^P_N)^{\dagger}(\sigma_y)\right)\equiv iS &=i
\left(\begin{array}{cccc}
0 & x_1 & x_2 & x_3\\
-x_1 & 0 & x_4 & x_5\\
-x_2 & -x_4 & 0 & x_6\\
-x_3 & -x_5 & -x_6 & 0
\end{array}\right), \nonumber
\eea
where $x_i$, $i=1,\dots,6$ are real free parameters and $s=2^{-N/2}$. The overall $8\times 8$ matrix, which needs to be positive is then given by
\bea
M(\tau_{P,N}) &= \left(\begin{array}{cc}
M(\eins_N) + M_z & M_x + S \\
M_x - S & M(\eins_N) - M_z
\end{array}\right)\\
&=\left(\begin{array}{cccccccc}
2 & 0 & 2s & 2s &                                           0 & \delta_Ns^2 + x_1 & s + x_2 & x_3 - s\\
0 & 0 & 0 & 0 &                                             \delta_Ns^2 - x_1 & 0 & x_4 + s & x_5 + (-1)^{N+1}s\\
2s & 0 & 1 & \delta_Ns^2 &                                  s - x_2 & s - x_4  & 1 & x_6\\
2s & 0 & \delta_Ns^2 & 1 &                                  -s - x_3 & (-1)^{N+1}s - x_5 & - x_6 & -1\\
0 & \delta_Ns^2 - x_1 & s - x_2 & - x_3 - s &               0 & 0 & 0 & 0\\
\delta_Ns^2 + x_1 & 0 & - x_4 + s & - x_5 + (-1)^{N+1}s &   0 & 2 & 2s & (-1)^N2s\\
s + x_2 & s + x_4 & 1 & - x_6 &                             0 & 2s & 1 & -\delta_Ns^2\\
-s + x_3 & (-1)^{N+1}s + x_5 & x_6 & -1 &                    0 & (-1)^N2s & -\delta_Ns^2 & 1
\end{array}\right),\nonumber
\eea
where $\delta_N=1-(-1)^N$.
\end{widetext}
The matrix $M(\tau_{P,N})$ is positive if and only if each of its principle minors is positive. This fixes all free parameters to be
\bea
x_1&=x_6=\delta_Ns^2\nonumber\\
x_3&=x_4=-s\label{BB84solutions}\\
x_2&=(-1)^Nx_5=s.\nonumber
\eea

\setcounter{thm}{0} \setcounter{equation}{0} \setcounter{figure}{0}
\renewcommand{\theequation}{D-\arabic{equation}}
\renewcommand{\thefigure}{D-\arabic{figure}}
\renewcommand{\thethm}{D-\arabic{thm}}

\section*{{\bf APPENDIX D}}
Here we provide some technical details used for the construction of the squashing model
of the active six-state measurement device for an intermediate postprocessing in Section \ref{ssection:6statealternativeCPP}.
In particular, we show the positivity of the matrix $\tau_{P,N,\textnormal{new}}(p)$
from Eq.~(\ref{eq:sixstatePOVMnewold}) for $p\geq1/3$.
First of all, for small integers $N$ we observe that the critical value of $p$ is equal to $1/3$.
For example, we need $p$ to exceed this value for $N=3$ in order matrix to be positive.
Then one can check directly that the matrix of $\tau_{P,N,\textnormal{new}}(p=1/3)$ is non-negative
for $N=1,2,...,10$. After this is done one can use the Gerschgorin disk theorem \cite{Gerschgorin31}
in order to prove that all eigenvalues strictly lie on the positive part of the real axis.
One can give an upper bound on each of the disk's (there are 12 of them) radii.
Each bound is a monotonically decreasing function of $N$ and for $N>10$ the union of 12 disks
lies on the part of the complex plain representing the eigenvalues of
the matrix, $\textnormal{Re}\lambda >0$. This implies the positivity of the eigenvalues since the matrix is Hermitian and all
its eigenvalues are real.


\setcounter{thm}{0} \setcounter{equation}{0} \setcounter{figure}{0}
\renewcommand{\theequation}{E-\arabic{equation}}
\renewcommand{\thefigure}{E-\arabic{figure}}
\renewcommand{\thethm}{E-\arabic{thm}}

\section*{{\bf APPENDIX E}}
\label{appendRuben}
Here we derive the POVM elements for the qudit measurement device from Section~\ref{rubenswork}. Formally, we will use $b^{\dagger}$ and $b$ to describe creation and annihilation operators of the output and $a^{\dagger}$ and $a$ to describe creation and annihilation operators of the input of the measurement device from Fig.~\ref{MUBapparatus}.
Assuming that the input state has $N$ photons, we have following POVM elements in terms of output operators
$b^{\dagger}$ and $b$:
\bea
F_N^{i,\alpha}&= \frac{1}{N!}\left(b^{\dagger}_{i,\alpha}\right)^N\ketbra{0}\left(b_{i,\alpha}\right)^N\nonumber\\
F_{\textnormal{mc},N}^{\alpha}&=\mathop{\sum_{m_k>0}}_{\sum m_k=N}
\bigotimes_{i=1}^d\frac{1}{m_k!}\left(b^{\dagger}_{i,\alpha}\right)^{m_k}\ketbra{0} \left(b_{i,\alpha}\right)^{m_k}\nonumber\\
F_{\textnormal{cc},N} &= \eins_N - \sum_{\alpha=0}^d\left(F_{\textnormal{mc},N}^{\alpha} + \sum_{i=1}^dF_N^{i,\alpha}\right)\label{POVMelements}.
\eea

Here $F_N^{i,\alpha}$ denotes a single click in detector $i$ of the detection module $\MM_{\alpha}$
(basis $\alpha$). $F_{\textnormal{mc},N}^{\alpha}$
denotes a multi-click for which several detectors of the same detection module $\MM_{\alpha}$
have clicked, whereas $F_{\textnormal{cc},N}$ denotes any cross-click between different detection modules.

In order to write the POVM elements in terms of input operators $a^{\dagger}$ and $a$ one needs to know the
input-output relations for the linear optical network. For an unbiased version ($p_{\alpha}=1/(d+1)$ for all $\alpha$)
of the beam splitter in Fig.~\ref{MUBapparatus}, the input and output modes are connected by the Fourier matrix:
\bea
&b_{\alpha,q}=\sum_{\beta=0}^d U_{\alpha,\beta}a_{\beta,q},\textnormal{ with }q=1\dots d,\\
&U_{\alpha,\beta}=(d+1)^{-\halbe}e^{\frac{2\pi i}{d+1}\alpha\beta}.\nonumber
\eea
Now we can substitute these relations in Eq.~(\ref{POVMelements}). We note that on the input side
only the mode $a_{\beta,q},\; q=1\dots d$ is occupied. Whence we can project the rest of the $a$'s on the
vacuum state. After some algebra the POVM elements in terms of input modes eventually become
\bea
F_N^{i,\alpha}&=(d+1)^{-N}\ketbraaa{N}{i,\alpha}{}\nonumber\\
F_{\textnormal{mc},N}^{\alpha}&=(d+1)^{-N}\left(\eins_{N}-\sum_{i}\ketbraaa{N}{i,\alpha}{}\right)\\
F_{\textnormal{cc},N} &= \left(1-\frac{1}{(d+1)^{N-1}}\right)\eins_N \nonumber.
\eea


\setcounter{thm}{0} \setcounter{equation}{0} \setcounter{figure}{0}
\renewcommand{\theequation}{F-\arabic{equation}}
\renewcommand{\thefigure}{F-\arabic{figure}}
\renewcommand{\thethm}{F-\arabic{thm}}

\section*{{\bf APPENDIX F}}
In this Appendix we provide the proof of
Proposition \ref{MUBPositivity} from Section \ref{rubenswork}.

{\it Proof of Proposition \ref{MUBPositivity}:}
The positivity is checked by proving that the Choi matrix of the map $\Lambda_{P,N}^{\dagger}$
is positive semi-definite. If $d$ is a prime number, the maximally entangled state can be decomposed
in the the chosen operator basis $\{Z_d^{\alpha},(X_dZ_d^{\alpha})^k\}_{\alpha,k}$,
$\alpha=0,\dots,d-1$, $k=1,\dots,d-1$ (one can
see this after some algebra involving Eqns.~(\ref{eq:MUBrelationship})) as
\bea
\ketbra{\psi^+}&=\frac{1}{d^2}\eins_d\otimes\eins_d + \frac{1}{d^2}\sum_{k=1}^{d-1}Z_d^{-k}\otimes Z_d^{k}\nonumber\\
&+\frac{1}{d^2}\sum_{\substack{\alpha=0\\ k=1}}^{d-1}\left(X_d Z_d^{-\alpha}\right)^k\otimes\left(X_d Z_d^{\alpha}\right)^{k}.\label{MUBMES}
\eea
Moreover, we point out that one can relate $\{Z_d^{\alpha},(X_dZ_d^{\alpha})^k\}_{\alpha,k}$
to the target POVM elements Eq.~(\ref{MUBtarget}):
\bea
Z_d^{k}&=(d+1)\sum_{i=0}^{d-1}\omega^{ik}F_1^{0,i},\nonumber\\
\left(X_d Z_d^{\alpha}\right)^{k}&= (d+1)\sum_{i=0}^{d-1}\omega^{ik}F_1^{\alpha+1,i}.\label{MUBObsPOVM}
\eea

We prove the positivity in two steps.

{\it Step 1 $N=1$:} When the QND measurement signals that one photon enters the measurement device,
the squashing map does not have to do anything and $\Lambda^{\dagger}=\eins$. Clearly it is completely
positive in this case.

{\it Step 2 $N\geq 2$:} When the QND measurement signals the presence of more than one photon in
the incoming signal the squashing map has to be non-trivial. The proof of its complete positivity
has several technical steps.

First of all, let us apply $\eins\otimes\Lambda_{P,N}^{\dagger}$ to the second term in Eq.~(\ref{MUBMES}).
We omit the overall factor of $1/d^2$ for brevity:
\bea
&\sum_{k=1}^{d-1}Z_d^{-k}\otimes \Lambda_{P,N}^{\dagger}\left[Z_d^{k}\right]=
\sum_{\substack{r,s=0\\k=1}}^{d-1}(d+1)^2\omega^{(s-r)k}\ftil_1^{r,0}\otimes\ftil_{P,N}^{s,0}\nonumber\\
&=(d+1)^2\left(d\sum_{r=0}^{d-1}\ftil_1^{r,0}\otimes\ftil_{P,N}^{r,0}-\sum_{r,s=0}^{d-1}\ftil_1^{r,0}\otimes\ftil_{P,N}^{s,0}\right),
\eea
where we used the first relation in Eq.~(\ref{MUBObsPOVM}) and the identity
$\sum_{k=1}^{d-1}\omega^{(s-r)k}=d\delta_{rs}-1$ for $\omega=\textnormal{e}^{\frac{2\pi i}{d}}$.
From the form of the basic POVM elements (Eq.~(\ref{MUBBasicPOVM}))
and the form of the full measurement POVM elements (Eq.~(\ref{PPRuben})) it follows that
\be
\sum_{r=0}^{d-1}\ftil_{P,N}^{r,\alpha} = \frac{1}{d+1}\eins_{P,N}
\forall \alpha\label{MUBhilfe}.
\ee
Note that this can also be concluded by a simple normalization argument: the POVM elements corresponding to clicks
in one of the detector modules should sum up to something proportional to the identity. The coefficient
of proportionality is equal to the probability $p_{\alpha}=1/(d+1)$ of the generalized balanced input beam splitter.

Hence
\bea
\sum_{k=1}^{d-1}Z_d^{-k}\otimes \Lambda_{P,N}^{\dagger}\left[Z_d^{k}\right]
&=d(d+1)^2\sum_{r=0}^{d-1}\ftil_1^{r,0}\otimes\ftil_{P,N}^{r,0} \nonumber\\
&-\eins_d\otimes\eins_{P,N}.\label{MUBSquashPos1}
\eea

Second, let us consider the action of $\eins\otimes\Lambda_{P,N}^{\dagger}$ on the third term in Eq.~(\ref{MUBMES}),
while omitting the overall factor of $1/d^2$ again for brevity.
\bea
&\sum_{\substack{\alpha=0\\ k=1}}^{d-1}\left(X_d Z_d^{-\alpha}\right)^k\otimes \Lambda_{P,N}^{\dagger}\left[\left(X_d Z_d^{\alpha}\right)^{k}
\right]\nonumber\\
&=(d+1)^2\sum_{\substack{r,s,\alpha=0\\k=1}}^{d-1}\omega^{(r+s)k}
\ftil_1^{r,-(\alpha+1)}\otimes\ftil_{P,N}^{s,\alpha+1}
\label{MUBSquashPos2}\\
&=d(d+1)^2\sum_{r,\alpha=0}^{d-1}
\ftil_1^{r,-(\alpha+1)}\otimes\ftil_{P,N}^{d-r,\alpha+1} \nonumber\\
&-d\eins_d\otimes\eins_{P,N}.\nonumber
\eea
Here we used the second relation in Eq.~(\ref{MUBObsPOVM}), the identity
$\sum_{k=1}^{d-1}\omega^{(r+s)k}=d\delta_{d,r+s}-1$ for $\omega=\textnormal{e}^{\frac{2\pi i}{d}}$, and
Eq.~(\ref{MUBhilfe}).

Putting Eq.~(\ref{MUBSquashPos1}) and Eq.~(\ref{MUBSquashPos2}) together we arrive at
\bea
&d^2\eins\otimes\Lambda_{P,N}^{\dagger}[\ketbra{\psi^+}]=d(d+1)^2\sum_{r=0}^{d-1}\ftil_1^{r,0}\otimes\ftil_{P,N}^{r,0} \nonumber\\
&+d(d+1)^2\sum_{r,\alpha=0}^{d-1}
\ftil_1^{r,-(\alpha+1)}\otimes\ftil_{P,N}^{d-r,\alpha+1}\label{MUBSquashPos3}\\
&-d\eins_d\otimes\eins_{P,N} \nonumber
\eea
The positivity of the last expression is proven by expanding the full measurement
POVM elements in terms of the basic POVM elements. For that,
we reformulate the first term in Eq.~(\ref{MUBSquashPos3}).
According to Eqns.~(\ref{MUBBasicPOVM})
and (\ref{MUBCPP}) for $\alpha=0$ we have
\bea
&d(d+1)^2\sum_{r=0}^{d-1}\ftil_1^{r,0}\otimes\ftil_{P,N}^{r,0}=
d(d+1)^2\sum_{r=0}^{d-1}\ftil_1^{r,0}\otimes\ftil_{\textnormal{any}}^{r,0}\nonumber\\
&+(d+1)\left(1-\frac{1}{(d+1)^{N-1}}\right)
\sum_{r=0}^{d-1}\ftil_1^{r,0}\otimes\eins_{P,N}\label{MUBhilfe1a}\\
&=d(d+1)^2\sum_{r=0}^{d-1}\ftil_1^{r,0}\otimes\ftil_{\textnormal{any}}^{r,0}
+\left(1-\frac{1}{(d+1)^{N-1}}\right)\eins_d\otimes\eins_{P,N}\nonumber,
\eea
where $\ftil_{\textnormal{any}}^{r,0}=F_{P,N}^{r,0}+F_{P,N}^{0}/d$ and we used Eq.~(\ref{MUBhilfe})
for the qudit part ($N=1$) of the tensor product.

For the second term in Eq.~(\ref{MUBSquashPos3}) we have
\bea
&d(d+1)^2\sum_{r,\alpha=0}^{d-1}
\ftil_1^{r,-(\alpha+1)}\otimes\ftil_{\textnormal{any}}^{d-r,\alpha+1}\nonumber\\
&=d(d+1)^2\sum_{r,\alpha=0}^{d-1}\frac{d}{p_{d-\alpha}p_{\alpha+1}}
\ftil_1^{r,-(\alpha+1)}\otimes\ftil_\textnormal{any}^{r,\alpha+1}\nonumber\\
&+(d+1)\left(1-\frac{1}{(d+1)^{N-1}}\right)\sum_{r,\alpha=0}^{d-1}
\ftil_1^{r,-(\alpha+1)}\otimes\eins_{P,N}\label{MUBhilfe1b}\\
&=d(d+1)^2\sum_{r,\alpha=0}^{d-1}
\ftil_1^{r,-(\alpha+1)}\otimes\ftil_{\textnormal{any}}^{r,\alpha+1}\nonumber\\
&+d\left(1-\frac{1}{(d+1)^{N-1}}\right)\eins_d\otimes\eins_{P,N}\nonumber,
\eea
where $\ftil_{\textnormal{any}}^{r,\alpha+1}=F_{P,N}^{d-r,\alpha+1}+F_{P,N}^{\alpha+1}/d$.

Substituting Eq.~(\ref{MUBhilfe1a}) and Eq.~(\ref{MUBhilfe1b}) into Eq.~(\ref{MUBSquashPos3}) yields
\bea
&d^2\eins\otimes\Lambda_P^{\dagger}[\ketbra{\psi^+}]\nonumber\\
&=d(d+1)^2\sum_{r=0}^{d-1}\left(\ftil_1^{r,0}\otimes\ftil_{\textnormal{any}}^{r,0}
+\sum_{\alpha=0}^{d-1}\ftil_1^{r,-(\alpha+1)}\otimes
\ftil_{\textnormal{any}}^{r,\alpha+1}\right)\label{MUBSquashPos4}\\
&+\left(1-\frac{1}{(d+1)^{N-2}}\right)\eins_d\otimes\eins_{P,N}\nonumber.
\eea
While the first term on the right hand side of the previous equation is strictly positive,
the eigenvalues of the second one are all equal to $1-\frac{1}{(d+1)^{N-2}}$ and are
non-negative for any $N\geq 2$. This finishes the proof of the second step and of the whole proposition.
\qed


\setcounter{thm}{0} \setcounter{equation}{0} \setcounter{figure}{0}
\renewcommand{\theequation}{G-\arabic{equation}}
\renewcommand{\thefigure}{G-\arabic{figure}}
\renewcommand{\thethm}{G-\arabic{thm}}

\section*{{\bf APPENDIX G}}
In this Appendix we provide the proof of Remark \ref{MUBCPPRemark}
from Section \ref{rubenswork}.

{\it Proof of Remark \ref{MUBCPPRemark}:}
first note that all basic POVM elements for unequal probabilities $p_{\alpha}$ generalize from
Eq.~(\ref{MUBBasicPOVM}) to
\bea
F_N^{i,\alpha}&=p_{\alpha}^N\ketbraaa{N}{i,\alpha}{}\nonumber\\
F_{\textnormal{mc},N}^{\alpha}&=p_{\alpha}^N\left(\eins_{N}-\sum_{i}\ketbraaa{N}{i,\alpha}{}\right)
\label{MUBdiffprobPOVM}\\
F_{\textnormal{cc},N} &= \left(1-\sum_{\alpha=0}^dp_{\alpha}^N\right)\eins_N \nonumber.
\eea

According to the postprocessing Eq.~(\ref{PPRuben}) the full measurement POVM elements then become
\bea
\ftil_N^{i,\alpha}&=p_{\alpha}^N\left(\ketbraaa{N}{i,\alpha}{}-
\frac{1}{d}\sum_{j=1}^d\ketbraaa{N}{j,\alpha}{}\right)\nonumber\\
&+\frac{(d+1)p_{\alpha}^N+1-\sum_{\beta=0}^{d}p_{\beta}^N}{d(d+1)}\eins_N.\label{MUBCPP}
\eea

Let us consider the map $\Lambda_{P_{\bot},N}$, which is applied when
an incoming $N$-photon state triggers the $P_{\bot}$ flag, i.e.~will with certainty produce a non-single-click. We assume that $\Lambda_{P_{\bot},N}$ fulfils the linear constraints in Eq.~(\ref{linconstrMUB}).
Then the probability of seeing a click in the $i$-th detector of the detection
module $\MM_{\alpha}$ is given by
\bea
&p(i,\alpha)=\tr{\vr_N^{\bot} \ftil_N^{i,\alpha}}
=\tr{\Lambda_{P_{\bot}}[\vr_N^{\bot}] \ftil_1^{i,\alpha}}\nonumber\\
&=\frac{\tr{\eins_D \ftil_1^{i,\alpha}}}{d}=\frac{1}{d}\label{MUBlinconstrcheck},
\eea
where we used the fact $\tr{A^{\dagger}B}=\tr{AB^{\dagger}}$.
The same probability can be re-expressed as
\be
p(i,\alpha)=
\tr{\vr_N^{\bot} P_{\bot}\ftil_N^{i,\alpha}P^{\dagger}_{\bot}}=\tr{\vr_N^{\bot} \ftil_{P_{\bot},N}^{i,\alpha}}.
\ee
By virtue of Eq.~(\ref{MUBCPP})
\be
\ftil_{P_{\bot},N}^{i,\alpha}
=\frac{(d+1)p_{\alpha}^N+1-\sum_{\beta=0}^{d}p_{\beta}^N}{d(d+1)}\eins_{P_{\bot},N},
\ee
and hence
\be
p(i,\alpha)=\frac{(d+1)p_{\alpha}^N+1-\sum_{\beta=0}^{d}p_{\beta}^N}{d(d+1)}.
\label{eq:probMUBcc}
\ee
Finally, a direct comparison of Eq.~(\ref{MUBlinconstrcheck}) and Eq.~(\ref{eq:probMUBcc}) implies that
\be
\frac{(d+1)p_{\alpha}^N+1-\sum_{\beta=0}^{d}p_{\beta}^N}{d(d+1)}=\frac{p_{\alpha}}{d}
\ee
must hold for any $N$ and for any $\alpha$.

Taking the limit $N\rightarrow \infty$ in the last equation we immediately see that the
only beam splitter ratio that respects the linear constraints is the one with
$p_{\alpha}=1/(d+1)$.
\qed

\end{document}